\pdfoutput=1

\documentclass[twocolumn]{article}

\usepackage[utf8]{inputenc}
\usepackage{amsmath,amssymb,amsfonts}
\usepackage{graphicx}
\usepackage{xcolor}
\usepackage[lined, boxed]{algorithm2e}
\usepackage{float}
\usepackage{listings}
\usepackage{booktabs}
\usepackage[justification=centering]{caption}
\usepackage{subcaption}
\usepackage[hidelinks]{hyperref}
\usepackage{pgfplots}
\usepackage{pgfplotstable}
\usepackage{tikz}
\usepackage{siunitx}
\usepackage{mathtools}

\usepackage[activate={true,nocompatibility},final,tracking=true,kerning=true,spacing=true,factor=1100,stretch=10,shrink=10]{microtype}
\microtypecontext{spacing=nonfrench}

\usepackage[backend=bibtex,bibstyle=authoryear,citestyle=authoryear,sorting=nyt,minnames=1,maxnames=3,nohashothers=true]{biblatex}
\DeclareBibliographyCategory{cited}
\AtEveryCitekey{\addtocategory{cited}{\thefield{entrykey}}}

\usetikzlibrary{plotmarks}
\usetikzlibrary{calc}

\newcommand{\NoSemiC}{\SetEndCharOfAlgoLine{\relax}}  
\newcommand{\DoSemiC}{\SetEndCharOfAlgoLine{\string;}}  

\let\frac\dfrac

\pgfplotsset{compat=1.16}

\title{Boosting Memory Access Locality of the Spectral Element Method with Hilbert Space-Filling Curves}

\author{
  Roger R. F. Ara\'ujo\textsuperscript{a},
  Lutz Gross\textsuperscript{b}
  and Samuel Xavier-de-Souza\textsuperscript{c} \\
  \\
  \small \textsuperscript{a} Programa de P\'os-Gradua\c{c}\~ao em Engenharia El\'etrica e de Computa\c{c}\~ao, \\
  \small Universidade Federal do Rio Grande do Norte, \\
  \small Natal/RN, Brazil (e-mail: roger\_rf\detokenize{@}dca.ufrn.br) \\
  \\
  \small \textsuperscript{b} School of Earth and Environmental Sciences, \\
  \small The University of Queensland, \\
  \small Brisbane, Queensland, Australia (e-mail: l.gross\detokenize{@}uq.edu.au) \\
  \\
  \small \textsuperscript{c} Departamento de Engenharia de Computa\c{c}\~ao e Automa\c{c}\~ao, \\
  \small Universidade Federal do Rio Grande do Norte, \\
  \small Natal/RN, Brazil (e-mail: samuel\detokenize{@}dca.ufrn.br) \\
  \\
  \small Corresponding author: \\
  \small Roger R. F. Ara\'ujo (e-mail: roger\_rf\detokenize{@}dca.ufrn.br)
  \normalsize
}

\date{April 2021}

\begin{document}

\maketitle

\begin{abstract}
We propose an algorithm based on Hilbert space-filling curves to reorder mesh elements in memory for use with the Spectral Element Method, aiming to attain fewer cache misses, better locality of data reference and faster execution.
We present a technique to numerically simulate acoustic wave propagation in 2D domains using the Spectral Element Method, and discuss computational performance aspects of this procedure. We reorder mesh-related data via Hilbert curves to achieve sizable reductions in execution time under several mesh configurations in shared-memory systems.
Our experiments show that the Hilbert curve approach works well with meshes of several granularities and also with small and large variations in element sizes, achieving reductions between 9\% and 25\% in execution time when compared to three other ordering schemes.


\end{abstract}

\section{Introduction}
\label{sec:introduction}

The efficient solution of the wave equation in terms of accuracy and computational performance is a topic of recurrent interest. The Spectral Element Method (SEM) \parencite{Patera1984,Kopriva2009}, a variant of the Finite Element Method (FEM) \parencite{Chaskalovic2008}, has been successfully applied to various wave-related problems \parencite{Zampieri2006,Bakir2012,Afanasiev2018}. SEM has proven to be versatile when it comes to the equations it can solve and to the geometry of the underlying domain, including complex media interfaces \parencite{Komatitsch1998}.

Both SEM and FEM discretize the domain as a mesh of interconnected elements to convert the target equation into a linear system in matrix form. FEM gives rise to sparse matrices, and the corresponding linear systems require specialized techniques to solve. In addition, the computational efforts to process those linear systems grow along with mesh resolution, posing a potential performance bottleneck. SEM avoids these complications by creating mass matrices that are diagonal by construction \parencite{Komatitsch1999} when using explicit time integration schemes. Diagonal mass matrices are desirable, as they minimize numerical diffusion and avoid matrix lumping. Moreover, the resulting linear systems are straightforward to solve and well-suited to parallel processing.

SEM-based software implementations need to traverse mesh elements and their nodes as they march an equation through time. Irregular and non-local memory accesses can severely harm this process. The layout of data structures, the sequence of memory accesses and cache hierarchy utilization all influence the efficiency of those traversals and overall execution performance \parencite{Sastry2014}. Improving spatial and temporal locality \parencite{Stallings2010} increases performance by avoiding cache misses.

In this work, we propose an approach based on Hilbert space-filling curves (SFCs) \parencite{Hilbert1891} to reorder mesh-related SEM data, mitigating irregular memory access and maximizing locality. We extended \textit{esys--escript} (Schaa et al., \cite*{SchaaGrossDuPlessis2016}), an open-source mathematical modeling tool, to support this approach and developed an SEM-based 2D acoustic wave simulator. We conducted experiments to compare the proposed Hilbert curve approach to different strategies using homogeneous and heterogeneous meshes and found that it works noticeably better, showing solid performance improvements in meshes of different granularities and large variations in element sizes. Finally, we show that our SEM formulation reliably simulates wave propagation in a complex, multilayered geological structure.

In the next section, we review prior work on SEM-based Partial Differential Equation (PDE) solvers and the use of SFCs to improve memory efficiency.
Section~\ref{sec:boosting_loc_sfc} describes how to generate standard Hilbert curves over squares, and a generalized approach that provides greater geometric flexibility.
In Section~\ref{sec:prop_method} we present our formulation to solve the wave equation using SEM with unstructured meshes of triangular elements and an explicit time integration scheme. In the same Section, we propose the algorithm to reorder mesh-related data using generalized Hilbert curves for memory-efficient traversals.
Section~\ref{sec:software_impl} discusses our implementation and optimization of the time integration scheme.
In Section~\ref{sec:results_and_discussion}, we conduct performance benchmarks of our proposed implementation with several memory reordering strategies and present a realistic wave propagation example. In the final section, we draw some conclusions.

\section{Related Work}
\label{sec:related_work}

Wave-type PDEs are essential in various fields such as acoustics, fluid dynamics and geophysics. Over time this resulted in the development of many wave equation solvers, of which we mention some SEM-based examples. \textcite{Komatitsch1998} solve elastic waves over unstructured meshes of quadrilaterals in 2D and hexahedra in 3D. \textcite{Komatitsch1999} discuss the simulation of seismic waves in 3D with unstructured meshes of hexahedral elements. Komatitsch et al. \parencite*{Komatitsch2000} use unstructured meshes of quadrilaterals in 2D and hexahedra in 3D to model wave propagation near a fluid-solid interface, taking into account the specific behavior of each medium. Mercerat et al. \parencite*{Mercerat2006} simulate elastic waves using unstructured meshes of triangles in 2D, in a work that employs SEM but does not use diagonal mass matrices. The SPECFEM3D Cartesian software package \parencite{Specfem3d} simulates several wave types over unstructured meshes of hexahedra. The \textit{p4est} software library (Burstedde et al., \cite*{Burstedde2011}), while not an equation solver per se, offers optimized data structures to work with unstructured meshes and uses forests of quadtrees in 2D and octrees in 3D for local refinement of elements. Our proposed implementation uses SEM to simulate acoustic wave propagation in 2D over unstructured meshes made of triangles.

Moving on to SFCs, they have the useful property of preserving spatial proximity between successive steps of the curve when marching through rectangles and hexahedra. This led to various applications in computer science that reorder memory or operations in the manner of SFC traversals to enhance cache utilization and memory efficiency. Voorhies uses Hilbert curves to reduce the number of paging operations required by computer screen scans when loading and discarding data related to objects randomly scattered across a screen \parencite{Arvo1991}. Mellor-Crummey et al. \parencite*{Mellor2001} employ SFCs to rearrange data and computations in particle problems, reducing cache misses and execution time. \textcite{Sastry2014} perform mesh warping through Laplace's equations, and use Hilbert curves to reorder mesh elements and vertices to achieve fewer cache misses and faster execution speed. The aforementioned \textit{p4est} library, whose main feature is parallel adaptive mesh refinement, is based on SFCs. We leverage a relatively recent formulation of generalized Hilbert curves to develop our algorithm to reorder mesh-related data.

\section{Boosting Locality with Space-Filling Curves}
\label{sec:boosting_loc_sfc}

According to the principle of locality of reference, when executing a program, a processor typically accesses the same memory areas repeatedly during short periods \parencite{Stallings2010}. A proper distribution of data among and within processors increases the probability that, after blocks from main memory get written to cache, memory references that the program makes in the near future are already available in the cache lines, helping to improve performance.

Upon first loading a mesh into memory from an external source such as a data file, we cannot expect its elements and nodes to follow any particular order. As a solver program marches a wave equation through time using SEM, it needs to traverse all mesh elements, and all local nodes of each element, at each step of its time integration scheme. If entities whose coordinates are close together in space lie far apart in memory, traversing them in the time integration scheme results in poor data locality. We can better exploit locality of reference by reordering data or computations to follow the traversal path of SFCs, thus leveraging their proximity-preserving properties. Potential improvements are better cache usage, memory efficiency and execution time.

\begin{figure}[!htb]
  \centering
  \resizebox{.95\linewidth}{!}{\includegraphics{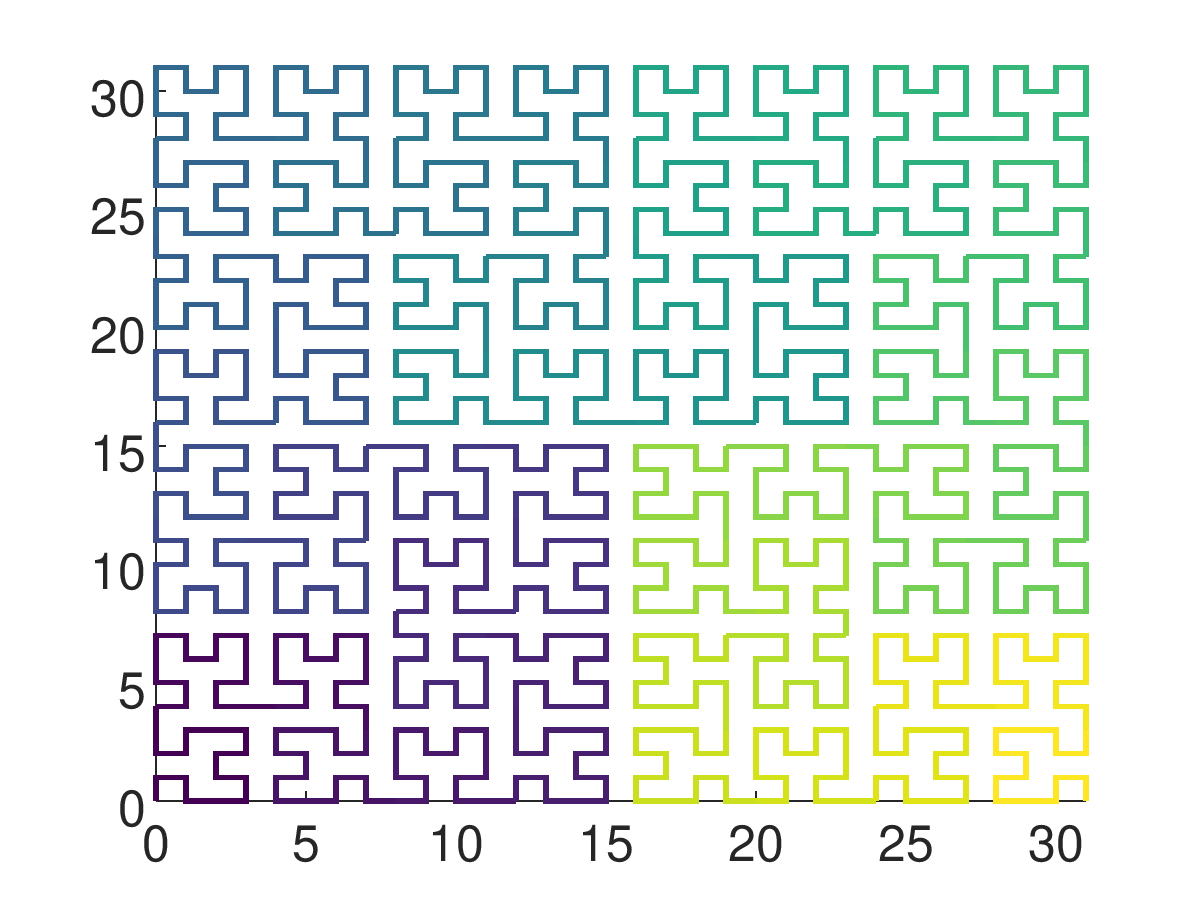}}
  \centering
  \caption{A standard Hilbert curve of order 5.}
  \label{fig:hilbert-curve}
\end{figure}

There are several SFC formulations we can use when developing memory reordering strategies, such as Peano curves \parencite{Peano1890}, Z-order curves \parencite{Morton1966} and Hilbert curves \parencite{Hilbert1891}. We focus our attention on Hilbert curves. Listing \ref{lst:hilbert} shows a simple, recursive Python generator of standard Hilbert curves, adapted from \textcite{Warren2013}. The curve runs through all integer $(x_0,x_1)$-coordinates inside the square of edge length $2^a - 1$, where $a$ is the order of the curve. The \textit{hilbert()} function continuously splits the square into smaller squares. The \textit{step()} function takes a step toward each smaller square, updating variables $x_0$, $x_1$ and $distance$. $x_0$ and $x_1$ are the intermediate coordinates along the traversal, and $distance$ is the accumulated distance. Figure \ref{fig:hilbert-curve} illustrates the traversal, displaying similar colors in the vicinity of intermediate steps.

\lstset{language=Python, frame=single}
\begin{lstlisting}[float=!htb, basicstyle=\small,
  captionpos=b, label={lst:hilbert},
  caption={Generator program for a standard Hilbert curve of order 5 (adapted from~\protect\cite{Warren2013}).}]
# Current coordinates within the square,
# distance traveled along the curve
x0, x1, distance = -1, 0, 0

# Values for direction (modulo 4):
#   0=right, 1=up, 2=left, 3=down
def step(direction):
    global x0, x1, distance
    direction = direction & 3
    if (direction == 0): x0 = x0 + 1
    if (direction == 1): x1 = x1 + 1
    if (direction == 2): x0 = x0 - 1
    if (direction == 3): x1 = x1 - 1
    print("%d %d" % (x0, x1))
    distance = distance + 1

# Values for rotation:
#   +1=clockwise, -1=counter-clockwise
def hilbert(direction, rotation, order):
    if (order == 0): return
    direction = direction + rotation
    hilbert(direction,
        -rotation, order - 1)
    step(direction)
    direction = direction - rotation
    hilbert(direction,
        rotation, order - 1)
    step(direction)
    hilbert(direction,
        rotation, order - 1)
    direction = direction - rotation
    step(direction)
    hilbert(direction,
        -rotation, order - 1)

order = 5
step(0)
hilbert(0, 1, order)
\end{lstlisting}

Listing \ref{lst:hilbert} exemplifies a typical practice in implementations of standard Hilbert curves, which is to use an order $a$ leading to a square of edge length $2^a - 1$ (i.e. a power of two). \textcite{Cerveny2020} proposes an alternative approach that constructs generalized Hilbert curves over rectangles with edges of arbitrary lengths, using the desired lengths as input. Its implementation, which again is recursive, results in a recursion depth that we observed to be greater than the order of the equivalent standard Hilbert curve. That said, as this alternative approach can deal with edge lengths that are not a power of two, the concepts of order and depth cannot be compared directly. The intermediate coordinates and accumulated distance that result from both approaches are equal when the edge lengths are a power of two.

\section{Proposed Method}
\label{sec:prop_method}

In this section we describe the wave equation, how we solve it using SEM, and an explicit time integration scheme for the numerical solution. We also present an algorithm to reorder mesh-related data according to the generalized Hilbert curves described in Section \ref{sec:boosting_loc_sfc}.

\subsection{The Wave Equation}

The acoustic wave equation in 2D (Feynman et al., \cite*{Feynman2011}) is a second-order PDE, that we restate as a first-order system over a domain $\Omega$:

\begin{equation}\label{eq:def_v_1}
\dot{v}_k = \frac{1}{\rho} u_{,k} \: ,
\end{equation}

\begin{equation}\label{eq:def_u_1}
\frac{1}{K} \dot{u} = v_{k,k} + f \: .
\end{equation}

The unknowns are $u$ (pressure) and $v_k$.  A single dot over a variable indicates a first time derivative. Variable $u$, a scalar, is continuous, whereas $v_k$, a vector with components for the $x_0$ and $x_1$ directions, is discontinuous. As for the remaining terms, $\rho$ is the density of the medium, $K$ is the compression modulus and $f$ is a source term. The values of $\rho$ and $K$ can vary with their location in the domain. The lower index $,k$ refers to the derivative with respect to direction $x_k$. We apply the Einstein convention with summation over double lower indices, i.e. in Eq. \eqref{eq:def_u_1} there is a summation over index $k$.

The initial conditions are:

\begin{equation}\label{eq:initial_cond}
u(\mathbf{x}, t = 0) = v_k(\mathbf{x}, t = 0) = 0 \: ,
\end{equation}

\noindent
for all locations $\mathbf{x}$ in the domain.
There may also be boundary conditions to consider, such as Dirichlet and Neumann-type conditions \parencite{Butcher2008} which are not shown here.

We can regard the acoustic wave equation as a particular case of a broader notation that describes wave-type PDEs in 2D and 3D. For a scalar, continuous wave field $u$ with a secondary, discontinuous field $v_k$, we express this as:

\begin{equation}\label{eq:def_u_2}
M \dot{u} = -(B_{kl} v_l)_{,k} + D u
+ \displaystyle\sum_{s} y^{(s)} (t) \delta_{\textbf{x}^{(s)}} \: ,
\end{equation}

\begin{equation}\label{eq:def_v_2}
\dot{v}_k = E_{kl} v_l + F_{kl} u_{,l} \: ,
\end{equation}

\noindent
where $M$, $B_{kl}$, $D$, $E_{kl}$ and $F_{kl}$ are PDE coefficients assumed to be constant over time but variable in space, and $y^{(s)}$ is a time-dependent wave source applied at point $\textbf{x}^{(s)}$. For the sake of a simpler presentation we assume that the continuous variable $u$ is a scalar and the discontinuous variable $v_k$ has two components. However, Eqs. \eqref{eq:def_u_2} and \eqref{eq:def_v_2} as well as the presented concepts can readily be extended to the more general case of a vector-valued continuous variable covering a wider range of applications such as elastic waves and Maxwell equations.  

For the acoustic wave equation in 2D restated in Eqs. \eqref{eq:def_u_1} and \eqref{eq:def_v_1}, $M$ and $D$ are scalar, whereas $B_{kl}$, $E_{kl}$ and $F_{kl}$ are matrices. This is $M = 1/K$, $B_{kk} = -1$ and $F_{kk} = 1/\rho$ for $k = 1, 2$. $D$ and $E_{kl}$ are not used and therefore contain zeros.

\subsection{The Spectral Element Method}
\label{sec:prop_method_sem}

The application of SEM requires converting the target PDEs into weak formulations, which for the generic Eq. \eqref{eq:def_u_2} we find to be:

\begin{equation}\label{eq:def_weak_u}
\begin{split}
\int_{\Omega} M \dot{u} q \: d\Omega =
& \int_{\Omega} B_{kl} v_l q_{,k} \: d\Omega \\
& + \int_{\Omega} D u q  \: d\Omega \\
& + \displaystyle\sum_{s} y^{(s)}(t) q(\textbf{x}^{(s)}) \: ,
\end{split}
\end{equation}

\noindent
where $q$ is an arbitrary test function that is smooth, in the sense that it is both continuous and piecewise differentiable.

We discretize Eqs. \eqref{eq:def_weak_u} and \eqref{eq:def_v_2} in space using SEM. We subdivide the domain $\Omega$ into an unstructured mesh of non-overlapping triangular elements ($\Omega^{(e)}$) covering the entire domain, where the upper index $e$ refers to the element count. All elements are based on a single reference element $\hat{\Omega}$ and described as $\textbf{x} = \mathcal{F}^{(e)}(\hat{\textbf{x}})$, with:

\begin{equation}
x_j = x^{(e)}_j + J^{(e)}_{j \hat{j}} \hat{x}_{\hat{j}}
\text{ for } \hat{\textbf{x}} = (\hat{x}_j) \in \hat{\Omega} \: ,
\end{equation}

\noindent
where $(x^{(e)}_j)$ is an offset point, and $(J^{(e)}_{j \hat{j}})$ is the Jacobian matrix describing the stretching and rotation of coordinates between $\hat{\Omega}$ and $\Omega^{(e)}$ for the transformation $\mathcal{F}^{(e)}$.

We represent the solution $u$ by its values $(U_\mu)$ at the global SEM nodes $(\textbf{x}_\mu)$ of the mesh. On each element, we approximate the solution $u$ by a polynomial of a given order defined by a local basis function $(\hat{N}_p)$ at the reference element $\hat{\Omega}$. On the reference element $\hat{\Omega}$, the local nodes $\hat{\textbf{x}}_q$ and local basis function $(\hat{N}_p)$ form a dual system:

\begin{equation}\label{eq:dual_sys}
\hat{N}_p(\hat{\textbf{x}}_q) = \delta_{pq}  \: .
\end{equation}{}

An important aspect of SEM is that the local nodes used to approximate the solution at each element are also quadrature points for numerical integration \parencite{Komatitsch1998}. In our proposed implementation, we use local nodes over variable-sized triangles and Appell polynomials as basis functions as described by \textcite{Blyth2005}. Resuming the previous discussion, we approximate the solution at $\textbf{x} \in \Omega^{(e)}$ as:

\begin{equation}
\begin{split}
u(\textbf{x}) = \displaystyle\sum_{p} \hat{N}_p(\hat{\textbf{x}}) U_{\mu(p,e)}
\text{ with } \textbf{x} = \mathcal{F}^{(e)} (\hat{\textbf{x}}) \: ,
\end{split}
\end{equation}{}

\noindent
where the index $\mu(p,e)$ maps the local SEM node $\hat{\textbf{x}}_p$ to the corresponding global node:

\begin{equation}
\textbf{x}_{\mu(p,e)} = \mathcal{F}^{(e)} (\hat{\textbf{x}}_p)  \: .
\end{equation}

Given the condition at Eq. \eqref{eq:dual_sys}, it follows that $U_\mu$ is the value of $u$ at global node $\textbf{x}_\mu$. The derivative of $u$ at element $\Omega^{(e)}$ is then given as:

\begin{equation}\label{eq:deriv_u}
\begin{split}
u_{,j}(\textbf{x}) = \displaystyle\sum_{p, \hat{j}}
  K^{(e)}_{j \hat{j}} \hat{N}_{p, \hat{j}}(\hat{\textbf{x}}) U_{\mu(p,e)} \\
\text{ for } \textbf{x} = \mathcal{F}^{(e)} (\hat{\textbf{x}}) \in \Omega^{(e)} \: ,
\end{split}
\end{equation}

\noindent
where $(K^{(e)}_{j \hat{j}})$ is the inverse matrix of $(J^{(e)}_{j \hat{j}})$.

We store the values of discontinuous function $v_k$ at the local SEM nodes $\hat{\textbf{x}}_p$ in element $e$ with values $V^{(e)}_{kp}$. Assuming that coefficients $E$ and $F$ are constant in each element with values $E^{(e)}$ and $F^{(e)}$, respectively, and applying the derivative $u_{,j}(\textbf{x})$ found at Eq. \eqref{eq:deriv_u}, we discretize Eq. \eqref{eq:def_v_2} at each local node in $\Omega^{(e)}$ as:

\begin{equation}\label{eq:v_change}
\begin{split}
\dot{V}^{(e)}_{kp} =
  & \displaystyle\sum_{l} E^{(e)}_{kl} V^{(e)}_{lp} \\
  & + \displaystyle\sum_{q, j, \hat{j}}
    F^{(e)}_{kj} K^{(e)}_{j \hat{j}} \hat{N}_{q, \hat{j}}(\hat{\textbf{x}}_p) U_{\mu (q,e)}
     \: .
\end{split}
\end{equation}

Proceeding to Eq. \eqref{eq:def_weak_u} we employ the fact that, with SEM, the local nodes of each element double as nodes for a numerical integration scheme. On the reference element, we express this as:

\begin{equation}\label{eq:numerical_integ}
\int_{\hat{\Omega}} \hat{f} \: d\hat{x} \approx
\displaystyle\sum_{q} \hat{\omega}_q \hat{f}(\hat{\textbf{x}}_q) \: ,
\end{equation}

\noindent
for any function $\hat{f}$ defined on $\hat{\Omega}$, where 
$\hat{\omega}_q$ are the integration weights. For a function $f$ defined on $\Omega$, we have:

\begin{equation}
\int_{\Omega} f \: dx =
\displaystyle\sum_e \int_{\Omega^{(e)}} f \: dx \approx
\displaystyle\sum_e \mathcal{J}^{(e)} \displaystyle\sum_q \hat{\omega}_q f_{qe} \: ,
\end{equation}

\noindent
with $f_{qe} = f(\textbf{x}_{\mu(q,e)})$ and $\mathcal{J}^{(e)} = det(J^{(e)}_{j \hat{j}} )$, where $\mathcal{J}^{(e)}$ is dependent on the element but independent of the integration node. Assuming that the PDE coefficients are constant on each element, we rewrite Eq. \eqref{eq:def_weak_u} as:

\begin{equation}\label{eq:u_change}
\dot{U}_{\nu} = \bar{M}_{\nu}^{-1} ( R_{\nu} + \bar{D}_{\nu} U_{\nu} + y_{\nu}(t)) \: ,
\end{equation}

\noindent
where for $A=M, D$:

\begin{equation}
\bar{A}_{\nu} = \displaystyle\sum_{e, q; \nu=\mu(q, e)}
  A^{(e)} \mathcal{J}^{(e)} \hat{\omega}_q  \: ,
\end{equation}
\noindent
and:
\begin{equation}\label{eq:u_change_r_effic}
R_{\nu} = \displaystyle\sum_{e, p, q; \nu=\mu(p, e)}
  \mathcal{J}^{(e)} \hat{\omega}_q
  \displaystyle\sum_{k, j, \hat{j}}
  B^{(e)}_{kj} V^{(e)}_{kq} K^{(e)}_{j \hat{j}}
  \hat{N}_{p, \hat{j}}(\hat{\textbf{x}}_q) ,
\end{equation}
\begin{equation}
y_{\nu}(t) = y^{(s)}(t)
  \text{ for } \textbf{x}^{(s)} = \textbf{x}_{\nu}  \: .
\end{equation}

Notice that calculating the rate of change for discontinuous variable $V^{(e)}_{kp}$ in Eq. \eqref{eq:v_change} can be performed independently for each element $e$, but requires the gather operation $U_{\mu (q,e)}$ to collect values of continuous variable $U_{\nu}$ at the global SEM nodes corresponding to the local SEM nodes at each element. In contrast, the rate of change of continuous variable $U_{\nu}$ in Eq. \eqref{eq:u_change} requires accumulating results of an element-by-element calculation at global SEM nodes. This operation creates a race condition when parallelized, as several elements can share global SEM nodes.

\subsubsection{Time Integration}
\label{sec:prop_method_time_integ}
The explicit time integration scheme starts at $n = 0$, with known $V^{(e, n=0)}_{kp} = 0$ and $U^{(n=0)}_{\nu} = 0$. Let $t^{(n)} = t^{(0)} + n \cdot h$, where $h$ is the time step size. With Eqs. \eqref{eq:u_change} and \eqref{eq:v_change} describing how $U$ and $V_k$ change over time, we use the Heun method \parencite{Heun1900} to update these variables at each time step. The predictor step is:

\begin{equation}
\begin{split}
\tilde{U}^{(n + 1)}_{\nu} & = U^{(n)}_{\nu} + h \cdot \dot{U}^{(n)}_{\nu} \: , \\
\tilde{V}^{(e, n+1)}_{kp} & = V^{(e, n)}_{kp} + h \cdot \dot{V}^{(e, n)}_{kp} \: ,
\end{split}
\end{equation}

\noindent
followed by the corrector step:

\begin{equation}
\begin{split}
U^{(n + 1)}_{\nu} & = U^{(n)}_{\nu}
  + \frac{h}{2} \cdot (\dot{U}^{(n)}_{\nu} + \dot{\tilde{U}}^{(n + 1)}_{\nu}) \: , \\
V^{(e, n+1)}_{kp} & = V^{(e, n)}_{kp}
  + \frac{h}{2} \cdot (\dot{V}^{(e, n)}_{kp} + \dot{\tilde{V}}^{(e, n + 1)}_{kp}) \: ,
\end{split}
\end{equation}

\noindent
where $\dot{\tilde{U}}^{(n + 1)}_{\nu}$ and $\dot{\tilde{V}}^{(e, n+1)}_{kp}$ are the results of Eqs. \eqref{eq:u_change} and \eqref{eq:v_change} evaluated using predictors $\tilde{U}^{(n + 1)}_{\nu}$ and $\tilde{V}^{(e, n+1)}_{kp}$ as input. We can use higher-order methods such as fourth-order Runge-Kutta \parencite{Runge1895,Kutta1901} for improved accuracy, at the cost of higher processing requirements.

\subsection{Memory Reordering with Generalized Hilbert Curves}
\label{sec:prop_method_hilbert}

To leverage the data locality provided by the spatial proximity features of generalized Hilbert curves, we propose the following memory reordering algorithm for SEM-based equation solvers:

\begin{enumerate}
\item Load the mesh into memory, and compute the centroids of all elements;

\item Let $w_m$ and $h_m$ be the width and the height of the mesh, $r$ the ratio $w_m / h_m$, and $n_e$ the number of elements. Define a bounding box with width $w_b = \sqrt{n_e}$ and height $h_b = \sqrt{n_e} / r$, made of subrectangles of width $w_s = w_b / w_m$ and height $h_s = h_b / h_m$. Using the technique proposed by \textcite{Cerveny2020}, generate a Hilbert curve over that bounding box, and store the $(x_{0},x_{1})$-coordinates of the intermediate steps of the curve;  

\item Let $(x_{m0},x_{m1})$ be the minimum coordinates of the mesh in the horizontal and vertical axes. Given the centroid of an element, let $(x_{c0},x_{c1})$ be its coordinates in the horizontal and vertical axes. Traverse the intermediate steps of the Hilbert curve stored previously, and, at each step, check which subrectangle of the bounding box the step is located in; check the elements of the mesh whose centroids map to that subrectangle (i.e. the elements for which the coordinates $(x_{c0} - x_{m0}) \cdot w_s$ and $(x_{c1} - x_{m1}) \cdot h_s$ are inside the subrectangle), and accumulate those elements in a list $L$. Each element must be stored in $L$ only once;

\item After traversing all elements, $L$ contains all elements and their order closely resembles that of the steps of the Hilbert curve. Now we must relabel elements and rearrange related data in memory, according to the optimized order contained in $L$.
\end{enumerate}

After executing this algorithm, sequential traversals of mesh elements closely match the steps of the Hilbert curve, resulting in increased data locality and memory efficiency. We apply this reordering to all data structures that are traversed element-by-element.

We employ the algorithm in the same way with structured or unstructured meshes. With structured meshes, we expect regular patterns in the number of elements found at intermediate steps of the Hilbert curve. With unstructured meshes, however, because of variable geometric complexity, we must not assume any uniformity in how many elements each step of the curve encompasses: the steps contained in certain regions may match many elements, whereas steps contained in other regions may match only a few elements (or possibly none), in an unpredictable fashion. An alternative to change the number of elements found at each step of the curve is to adjust the width and height of the bounding box while honoring the proportion $w_b / h_b = r$ (for instance, using $w_b = \sqrt{n_e} \cdot r$ and $h_b = \sqrt{n_e}$).

We can achieve additional memory efficiency
by also reordering global nodes. The proposed procedure, based on the Cuthill--McKee algorithm \parencite{Cuthill1969}, follows:

\begin{enumerate}
\item Traverse all elements; at each element, accumulate all of its nodes in a list $N^{(e)}$. A node must not be stored in $N^{(e)}$ if it has been processed in a previous element;

\item Each node in $N^{(e)}$ has a degree, i.e. the number of connections that the node has to other nodes. Sort the nodes in $N^{(e)}$ by ascending degree; relabel the nodes and rearrange related data in memory, according to the optimized order contained in $N^{(e)}$.
\end{enumerate}

We apply this reordering strategy to all data structures that are traversed node-by-node.

\section{Software Implementation}
\label{sec:software_impl}

We developed a 2D acoustic wave propagation simulator in the \textit{esys--escript} tool, using the concepts and equations discussed in Section~\ref{sec:prop_method_sem} and the Heun scheme described in Section~\ref{sec:prop_method_time_integ}. We built its central functionality --- the explicit time integration scheme --- in C++, leveraging OpenMP \parencite{Openmp2021} for multithreading.

In describing the implementation of the time integration scheme, we assess the computational effort of traversing data structures element-by-element at each time step. To that end, we analyze Eqs. \eqref{eq:v_change} and \eqref{eq:u_change}, which describe how variables $U$ and $V_k$ change over time. The layout of their terms in memory (see Figure \ref{fig:mem_layout}) is important regarding the behavior of the processor cache:

\begin{itemize}
\item PDE coefficients $\bar{M}_{\nu}$ and $\bar{D}_{\nu}$ and continuous variable $U_{\nu}$ are arrays of scalars, with each item in the arrays corresponding to a global node $\nu$. We store them in the order of continuous labeling of the global nodes;

\item We define PDE coefficients $B^{(e)}_{kl}$, $E^{(e)}_{kl}$ and $F^{(e)}_{kl}$ at each element $e$ as matrices, where $k$ is the row and $l$ is the column. We store them in the order of continuous labeling of the elements, following a Fortran standard storing rows continuously in memory;

\item We define discontinuous variable $V^{(e)}_{kp}$ at each local node $p$ of each element $e$, with two components for directions $x_0$ and $x_1$ indexed by $k$. We store its instances in the order of continuous labeling of the elements.
\end{itemize}

\newcounter{nodeidx}
\setcounter{nodeidx}{1}
\newcommand{\nodes}[1]{%
  \foreach \num in {#1}{
    \node[minimum height=8mm, minimum width=8.5mm, draw, rectangle] (\arabic{nodeidx}) at (\arabic{nodeidx}*0.9,0) {\num};
    \stepcounter{nodeidx}
  }
}

\newcommand{\brckt}[4]{
  \draw
  (#1.south west)
  ++($(-.0, -.1)
  + (-.0*#3, 0)$)
  --
  ++($(0, -.1)
  + (0, -#3*1.25em)$)
  --
  node [below] {#4}
  ($(#2.south east)
  + (.0, -.1)
  + (.0*#3, 0)
  + (0, -.1)
  + (0, -#3*1.25em)$)
  --
  ++($(0, #3*1.25em)
  + (0, .1)$);%
}

\begin{figure}[!htb]
\noindent
\centering
\small
\setcounter{nodeidx}{1}
\begin{tikzpicture}
  \nodes{$\bar{M}_1$}
  \nodes{$\bar{M}_2$}
  \nodes{$\dotsm$}
  \brckt{1}{3}{0}{global nodes}
\end{tikzpicture}
\setcounter{nodeidx}{1}
\begin{tikzpicture}
  \nodes{$B^{(1)}_{1,1}$,$B^{(1)}_{1,2}$,$B^{(1)}_{2,1}$,$B^{(1)}_{2,2}$}
  \nodes{$B^{(2)}_{1,1}$,$B^{(2)}_{1,2}$,$B^{(2)}_{2,1}$,$B^{(2)}_{2,2}$}
  \nodes{$\dotsm$}
  \brckt{1}{4}{0}{element 1}
  \brckt{5}{8}{0}{element 2}
\end{tikzpicture}
\setcounter{nodeidx}{1}
\begin{tikzpicture}
  \nodes{$V^{(1)}_{1,1}$,$V^{(1)}_{2,1}$}
  \nodes{$\dotsm$}
  \nodes{$V^{(1)}_{1,p}$,$V^{(1)}_{2,p}$}
  \nodes{$V^{(2)}_{1,1}$,$V^{(2)}_{2,1}$}
  \nodes{$\dotsm$}
  \nodes{$\dotsm$}
  \brckt{1}{2}{0}{\footnotesize local node 1 \small}
  \brckt{4}{5}{0}{\footnotesize local node $p$ \small}
  \brckt{1}{5}{1}{element 1}
  \brckt{6}{8}{1}{element 2}
\end{tikzpicture}
\normalsize
\caption{Memory layout of equation terms for global node-based values $\bar{M}_{\nu}$, element-based values $B^{(e)}_{kl}$ and local node-based values $V^{(e)}_{kp}$. $\bar{D}_{\nu}$ and $U_{\nu}$ have the same layout as $\bar{M}_{\nu}$, and $E^{(e)}_{kl}$ and $F^{(e)}_{kl}$ have the same layout as $B^{(e)}_{kl}$.}
\label{fig:mem_layout}
\end{figure}
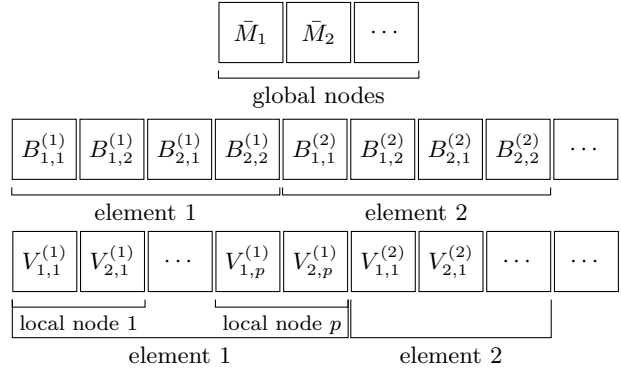

We examine Eq. \eqref{eq:v_change} for the change in $V_k$. The pseudo-code for its first term, $\sum_{l} E^{(e)}_{kl} V^{(e)}_{lp}$, is in the first code block of Algorithm \ref{alg:v_change}. The main factors for the computational complexity of this term are the number of elements and the number of nodes per element, which stems from the polynomial approximation order.
The remainder of this discussion considers the case where the discontinuous variable $v_k$ has two components, therefore $k$ varies from $1$ to $2$. Since $E^{(e)}_{kl}$ is a square matrix in this case, $l$ behaves the same as $k$ and also varies from $1$ to $2$. This assumption of $k$ and $l$ allows to calculate the initial term of $\dot{V}^{(e)}_{kp}$ in a single statement without looping over $k$ and $l$.

\begin{algorithm}[!htb]
\ForEach{element $e$}{
  \tcp{First term}
  \ForEach{node $p$ in $e$}{
    $\dot{V}^{(e)}_{1,p} \leftarrow E^{(e)}_{1,1} V^{(e)}_{1,p}
      + E^{(e)}_{1,2} V^{(e)}_{2,p} $\;
    $\dot{V}^{(e)}_{2,p} \leftarrow E^{(e)}_{2,1} V^{(e)}_{1,p}
      + E^{(e)}_{2,2} V^{(e)}_{2,p} $\;
  }
  \NoSemiC \; \DoSemiC
  \tcp{Second term}
  let $FK[1..2][1..4]$ : array of real\;
  \For{component $k$ in $1, 2$}{
    $FK[k][1] \leftarrow F^{(e)}_{k,1} K^{(e)}_{1,1}$\;
    $FK[k][2] \leftarrow F^{(e)}_{k,2} K^{(e)}_{2,1}$\;
    $FK[k][3] \leftarrow F^{(e)}_{k,1} K^{(e)}_{1,2}$\;
    $FK[k][4] \leftarrow F^{(e)}_{k,2} K^{(e)}_{2,2}$\;
  }
  \NoSemiC \; \DoSemiC
  let $U_2[1..q]$ : array of real\;
  \ForEach{integration point $q$ in $e$}{
    $U_2[q] \leftarrow U_{\mu(q,e)}$\;
  }
  \NoSemiC \; \DoSemiC
  \For{component $k$ in $1, 2$}{
    \ForEach{node $p$ in $e$}{
      \ForEach{integration point $q$ in $e$}{
        $\dot{V}^{(e)}_{kp} \leftarrow \dot{V}^{(e)}_{kp}$ \\
        $\: + FK[k][1]
          \hat{N}_{q,1}(\hat{\textbf{x}}_p)
          U_2[q]$  \\
        $\: + FK[k][2]
          \hat{N}_{q,1}(\hat{\textbf{x}}_p)
          U_2[q]$ \\
        $\: + FK[k][3]
          \hat{N}_{q,2}(\hat{\textbf{x}}_p)
          U_2[q]$ \\
        $\: + FK[k][4]
          \hat{N}_{q,2}(\hat{\textbf{x}}_p)
          U_2[q]$\;
      }
    }
  }
}
\caption{Calculation of $\dot{V}^{(e)}_{kp}$ by Eq. \eqref{eq:v_change} for a discontinuous variable $v_k$ with two components.}
\label{alg:v_change}
\end{algorithm}

The second term, a summation over $q$, $j$ and $\hat{j}$, is more expensive than the first. As shown in the pseudo-code, the $\dot{V}^{(e)}_{kp}$ accumulation step requires three nested loops where we introduce two precalculations to eliminate redundant calculations. For the first precalculation of array $FK$, as $k$ varies from $1$ to $2$ and $F^{(e)}_{kj}$ and $K^{(e)}_{j \hat{j}}$ are square matrices, $j$ and $\hat{j}$ vary from $1$ to $2$. Assuming $F^{(e)}_{kj}$ to be equal for all nodes within an element, we compute the values of $F^{(e)}_{kj} K^{(e)}_{j \hat{j}}$ in a single statement eliminating iterations over directions $j$ and $\hat{j}$; we store them in the $FK$ array, which is independent of local node $p$ and integration point $q$ and reduces memory accesses and arithmetic operations in the accumulation step later. Moving on to the second precalculation, we need to access the value of solution $(U_\mu)$ at global node $\mu({q,e})$, a gather operation. We cache the values of $U_{\mu(q,e)}$ in the $U_2$ array to improve data locality and reduce potential repeated cache line fetches. Finally, at the $\dot{V}^{(e)}_{kp}$ accumulation step, the value of $q$ is equal to the number of nodes per element; given the assumptions for $j$ and $\hat{j}$, we write the accumulation expression in a single statement without looping over $j$ and $\hat{j}$. Since both terms of Eq. \eqref{eq:v_change} do not affect data of elements other than the current element, Algorithm \ref{alg:v_change} is straightforward to parallelize by distributing elements among separate threads.

Proceeding to Eq. \eqref{eq:u_change}, which describes how $U$ changes over time, we start by examining its second term, $\bar{M}^{-1}_\nu \bar{D}_\nu U_\nu$, in Algorithm \ref{alg:u_change_1}. We precalculate $Y_{\nu} =\bar{M}^{-1}_\nu \bar{D}_\nu$ for all global nodes before time integration and use it to initialize $\dot{U}_{\nu}$. This term of Eq. \eqref{eq:u_change} is straightforward to parallelize, as it does not affect the data of global nodes other than the current node.

\begin{algorithm}[!htb]
\small
\tcp{Second term}
\ForEach{global node $\nu$}{
  $\dot{U}_{\nu} \leftarrow Y_{\nu} U_{\nu}$\;
}
\NoSemiC \; \DoSemiC
\tcp{First term}
\ForEach{color $c$}{
  \ForEach{element $e$}{
    \If{color of element $e \neq c$} {
      go to next element\;
    }
    \NoSemiC \; \DoSemiC
    let $BK[1..2][1..4]$ : array of real\;
    \For{$k$ in $1, 2$}{
      $BK[k][1] \leftarrow B^{(e)}_{k,1} K^{(e)}_{1,1}$\;
      $BK[k][2] \leftarrow B^{(e)}_{k,2} K^{(e)}_{2,1}$\;
      $BK[k][3] \leftarrow B^{(e)}_{k,1} K^{(e)}_{1,2}$\;
      $BK[k][4] \leftarrow B^{(e)}_{k,2} K^{(e)}_{2,2}$\;
    }
    \NoSemiC \; \DoSemiC
    \ForEach{local node $p$ in $e$}{
      $aux_1 \leftarrow \bar{M}^{-1}_{\nu=\mu(p,e)} \mathcal{J}^{(e)}$\;
      $val \leftarrow 0$\;
      \NoSemiC \; \DoSemiC
      \For{$k$ in $1, 2$}{
        \ForEach{integration point $q$ in $e$}{
          $aux_2 \leftarrow aux_1 \cdot \hat{\omega}_q V^{(e)}_{kq}$\;
          $val \leftarrow val$ \\
          $\: + BK[k][1] \cdot aux_2 \cdot
            \hat{N}_{p,1}(\hat{\textbf{x}}_q)$ \\
          $\: + BK[k][2] \cdot aux_2 \cdot
            \hat{N}_{p,1}(\hat{\textbf{x}}_q)$ \\
          $\: + BK[k][3] \cdot aux_2 \cdot
            \hat{N}_{p,2}(\hat{\textbf{x}}_q)$ \\
          $\: + BK[k][4] \cdot aux_2 \cdot
            \hat{N}_{p,2}(\hat{\textbf{x}}_q)$\;
        }
      }
      \NoSemiC \; \DoSemiC
      $\dot{U}_{\mu(p,e)} \leftarrow \dot{U}_{\mu(p,e)} + val$\;
    }
  }
}
\normalsize
\caption{Calculation of $\dot{U}_{\nu}$ via the second and first terms of Eq. \eqref{eq:u_change} with precalculated $Y_{\nu}=\bar{M}^{-1}_\nu \bar{D}_{\nu}$.}
\label{alg:u_change_1}
\end{algorithm}

The first term of Eq. \eqref{eq:u_change}, $\bar{M}^{-1}_\nu R_\nu$, involves more computational work. We obtain $R_\nu$ by summation over all elements $e$, local nodes $p$ and integration points $q$ with $\nu=\mu(p, e)$. We must loop over the elements and their local nodes, adding contributions to the respective global nodes. Analogously to $F^{(e)}_{kj} K^{(e)}_{j \hat{j}}$ in Algorithm \ref{alg:v_change}, there is a precalculation section where we compute the values of $B^{(e)}_{kj} K^{(e)}_{j \hat{j}}$; we store them in the $BK$ array, which is independent of $p$ and $q$ and eliminates redundant arithmetic operations in the accumulation step later.

The accumulation of contributions to add to $\dot{U}_{\nu=\mu(p,e)}$ occurs in the $val$ variable at the nested for-loops that traverse $k$ and $q$, and we perform it in a single statement without loops, exploiting the assumptions on $j$ and $\hat{j}$ counts. The update of $\dot{U}_{\nu}$ with the accumulated value stored at $val$ is a gather operation, as the solution value for global node $\nu = \mu(p,e)$ receives contributions of all local nodes within elements that map to that global node, and a global node may be shared among neighboring elements.

Since local nodes of several elements may share a single global node, we must be careful when updating $\dot{U}_{\nu}$ and processing multiple elements simultaneously. If any two elements that share a global node $\nu$ update $\dot{U}_{\nu}$ concurrently, race conditions may occur. To avoid that situation, we use element coloring (Davies et al., \cite*{Davies2004}). When we first construct the mesh, we regard the elements of the mesh as vertices of a graph. We consider any two elements sharing a global node to be connected by an edge in the graph, and use a greedy vertex-coloring algorithm to assign different colors to any two elements sharing a global node. This allows processing elements of identical color simultaneously without mishandling operations over $\dot{U}_{\nu}$. That way, whereas we must traverse the available colors sequentially in the $c$ for-loop of Algorithm \ref{alg:u_change_1}, we can safely parallelize the remaining work at the $e$ for-loop that follows. The number of colors is contingent on mesh complexity.

\begin{algorithm}[!t]
\ForEach{wave source $s$}{
  $\nu \leftarrow$ global node that $\textbf{x}^{(s)}$ maps to\;
  $\dot{U}_{\nu} \leftarrow \dot{U}_{\nu} + (\bar{M}^{-1}_{\nu} y_{\nu}(t))$\;
}
\caption{Calculation of the source term, which is the third term of Eq. \eqref{eq:u_change}.}
\label{alg:u_change_2}
\end{algorithm}

The third and final term in Eq.~\eqref{eq:u_change}, $\bar{M}^{-1}_{\nu} y_{\nu}(t)$, is simple. As listed in Algorithm \ref{alg:u_change_2}, we need only to traverse the global nodes in the domain that contain wave sources and accumulate their contributions into $\dot{U}_{\nu}$. The accumulation step for $\dot{U}_{\nu}$ assumes that Algorithm \ref{alg:u_change_1} was executed beforehand.
Typically the number of point sources is small and parallelizing this segment is not worthwhile.

We need to execute the steps in Algorithms \ref{alg:v_change} through \ref{alg:u_change_2} to compute Eqs. \eqref{eq:v_change} and \eqref{eq:u_change} at every step of the time integration scheme. As per Heun's method, we compute those equations twice at every time step, first for the predictor phase and then for the corrector phase.

\section{Results and Discussion}
\label{sec:results_and_discussion}

In this section, we assess the execution speed of our proposed implementation under a number of scenarios. We follow by demonstrating the application of our acoustic wave simulator to a complex, realistic domain.

\subsection{Performance Benchmarks}
\label{sec:benchmarks}

We executed two benchmarks of our simulator using meshes with linear elements of synthetic domains generated with the mesh generator tool Gmsh \parencite{Gmsh2009}. As Gmsh does not generate SEM meshes, we added SEM nodes to the meshes in memory using the locations described by ~\textcite{Blyth2005}. The simulator supports four memory reordering strategies for mesh-related data, described next.

\medskip

\noindent
\textbf{No Strategy}:
In this strategy, we simply store element and node-related data in the same order as obtained from the original source. Generally speaking, we cannot make any assumptions on the data order. However, this should not necessarily be considered as a random ordering.

\medskip

\noindent
\textbf{Node Connectivity Strategy}:
This strategy is very similar to the Cuthill--McKee algorithm \parencite{Cuthill1969}. We pick the node with the smallest number of connections in the mesh and proceed to the node with the smallest number of connections that is connected to that node; we repeat the process, avoiding previously visited nodes, until all nodes have been traversed. At each node, we accumulate the element that the node belongs to in a list and the node itself in another list, avoiding duplicate items. Finally, we relabel the elements and nodes and reorder related data in memory following the order contained in the lists.

\medskip

\noindent
\textbf{Node Distance Strategy}:
This strategy closely resembles the node connectivity strategy, the only difference being that we pick the node closest to a reference point and proceed to the node closest to the reference point that is connected to that node. We repeat this process, avoiding previously visited nodes, until all nodes have been traversed. In our experiments, we always positioned the chosen reference point in the lower-left corner of the domain.

\medskip

\noindent
\textbf{Hilbert SFC Strategy}:
In this strategy, we apply the algorithm described in Section \ref{sec:prop_method_hilbert}.

\medskip

In our benchmarks, after loading meshes from local files and adding SEM nodes, memory reordering made the mesh preparation process take between 15\% and 20\% longer than performing no reordering. In the following discussions, the labels ``None'', ``Conn.'', ``Dist.'' and ``SFC'' refer to the application of no strategy, node connectivity, node distance and Hilbert curve memory reordering strategy, respectively.

\subsubsection{Single-Layered Domain}

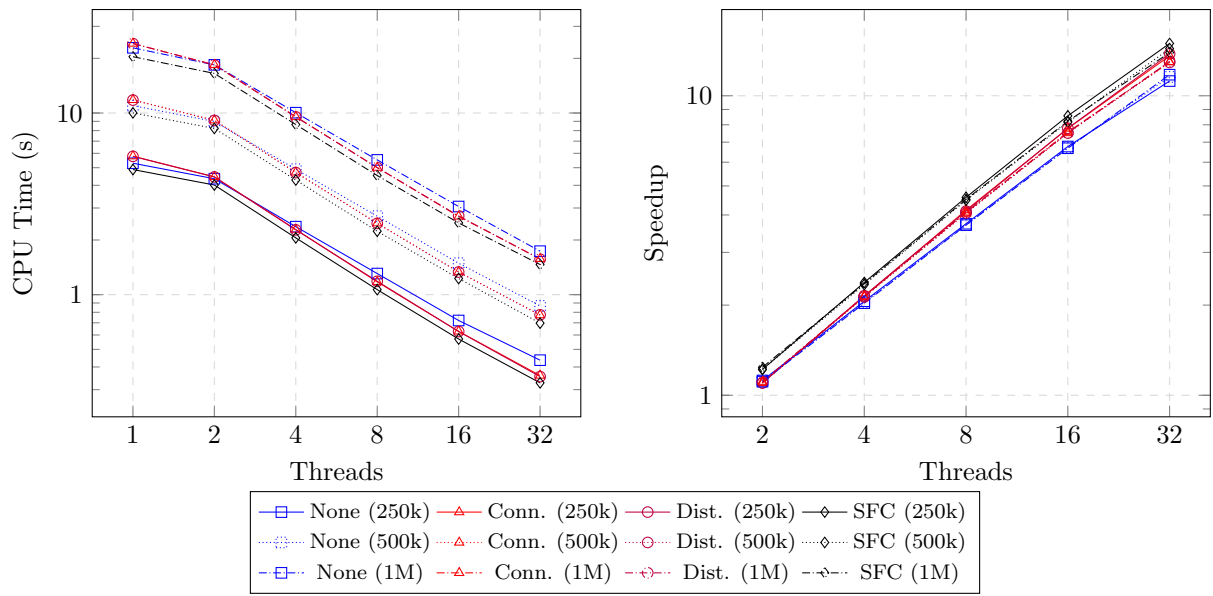
\begin{figure*}[!htb]
\noindent
\centering
\begin{minipage}[b]{0.49\linewidth}
\begin{tikzpicture}
\begin{loglogaxis}[
  height=7.0cm,
  width=\linewidth,
  grid=major, grid style={dashed,gray!30},
  xlabel=Threads, ylabel=CPU Time (s),
  xtick={ 1, 2, 4, 8, 16, 32 },
  log ticks with fixed point,
  legend columns=4,
  legend to name=leg:avg_cpu_time_threads_exp_1
  ]
  \addplot+[color=blue,mark=square,style=solid]
  table[x=c1,y=c2,col sep=semicolon] {anc/avg-cpu-time-thr-exp-1-250k.csv};
  \addlegendentry{\footnotesize  None (250k)}
  \addplot+[color=red,mark=triangle,style=solid]
  table[x=c1,y=c3,col sep=semicolon] {anc/avg-cpu-time-thr-exp-1-250k.csv};
  \addlegendentry{\footnotesize  Conn. (250k)}
  \addplot+[color=purple,mark=o,style=solid]
  table[x=c1,y=c4,col sep=semicolon] {anc/avg-cpu-time-thr-exp-1-250k.csv};
  \addlegendentry{\footnotesize  Dist. (250k)}
  \addplot+[color=black,mark=diamond,style=solid]
  table[x=c1,y=c5,col sep=semicolon] {anc/avg-cpu-time-thr-exp-1-250k.csv};
  \addlegendentry{\footnotesize  SFC (250k)}

  \addplot+[color=blue,mark=square,style=densely dotted] 
  table[x=c1,y=c2,col sep=semicolon] {anc/avg-cpu-time-thr-exp-1-500k.csv};
  \addlegendentry{\footnotesize None (500k)}
  \addplot+[color=red,mark=triangle,style=densely dotted] 
  table[x=c1,y=c3,col sep=semicolon] {anc/avg-cpu-time-thr-exp-1-500k.csv};
  \addlegendentry{\footnotesize Conn. (500k)}
  \addplot+[color=purple,mark=o,style=densely dotted] 
  table[x=c1,y=c4,col sep=semicolon] {anc/avg-cpu-time-thr-exp-1-500k.csv};
  \addlegendentry{\footnotesize Dist. (500k)}
  \addplot+[color=black,mark=diamond,style=densely dotted] 
  table[x=c1,y=c5,col sep=semicolon] {anc/avg-cpu-time-thr-exp-1-500k.csv};
  \addlegendentry{\footnotesize SFC (500k)}

  \addplot+[color=blue,mark=square,style=densely dashdotted] 
  table[x=c1,y=c2,col sep=semicolon] {anc/avg-cpu-time-thr-exp-1-1m.csv};
  \addlegendentry{\footnotesize None (1M)}
  \addplot+[color=red,mark=triangle,style=densely dashdotted] 
  table[x=c1,y=c3,col sep=semicolon] {anc/avg-cpu-time-thr-exp-1-1m.csv};
  \addlegendentry{\footnotesize Conn. (1M)}
  \addplot+[color=purple,mark=o,style=densely dashdotted] 
  table[x=c1,y=c4,col sep=semicolon] {anc/avg-cpu-time-thr-exp-1-1m.csv};
  \addlegendentry{\footnotesize Dist. (1M)}
  \addplot+[color=black,mark=diamond,style=densely dashdotted] 
  table[x=c1,y=c5,col sep=semicolon] {anc/avg-cpu-time-thr-exp-1-1m.csv};
  \addlegendentry{\footnotesize SFC (1M)}
\end{loglogaxis}
\end{tikzpicture}
\end{minipage}
\begin{minipage}[b]{0.49\linewidth}
\centering
\begin{tikzpicture}
\begin{loglogaxis}[
  height=7.0cm,
  width=\linewidth,
  grid=major, grid style={dashed,gray!30},
  xlabel=Threads, ylabel=Speedup,
  legend columns=2, legend pos=north west,
  xtick={ 1, 2, 4, 8, 16, 32 },
  log ticks with fixed point
  ]
  \addplot+[color=blue,mark=square,style=solid]
  table[x=c1,y=c2,col sep=semicolon] {anc/avg-speedup-thr-exp-1-250k.csv};
  \addplot+[color=red,mark=triangle,style=solid]
  table[x=c1,y=c3,col sep=semicolon] {anc/avg-speedup-thr-exp-1-250k.csv};
  \addplot+[color=purple,mark=o,style=solid]
  table[x=c1,y=c4,col sep=semicolon] {anc/avg-speedup-thr-exp-1-250k.csv};
  \addplot+[color=black,mark=diamond,style=solid]
  table[x=c1,y=c5,col sep=semicolon] {anc/avg-speedup-thr-exp-1-250k.csv};

  \addplot+[color=blue,mark=square,style=densely dotted] 
  table[x=c1,y=c2,col sep=semicolon] {anc/avg-speedup-thr-exp-1-500k.csv};
  \addplot+[color=red,mark=triangle,style=densely dotted] 
  table[x=c1,y=c3,col sep=semicolon] {anc/avg-speedup-thr-exp-1-500k.csv};
  \addplot+[color=purple,mark=o,style=densely dotted] 
  table[x=c1,y=c4,col sep=semicolon] {anc/avg-speedup-thr-exp-1-500k.csv};
  \addplot+[color=black,mark=diamond,style=densely dotted] 
  table[x=c1,y=c5,col sep=semicolon] {anc/avg-speedup-thr-exp-1-500k.csv};

  \addplot+[color=blue,mark=square,style=densely dashdotted] 
  table[x=c1,y=c2,col sep=semicolon] {anc/avg-speedup-thr-exp-1-1m.csv};
  \addplot+[color=red,mark=triangle,style=densely dashdotted] 
  table[x=c1,y=c3,col sep=semicolon] {anc/avg-speedup-thr-exp-1-1m.csv};
  \addplot+[color=purple,mark=o,style=densely dashdotted] 
  table[x=c1,y=c4,col sep=semicolon] {anc/avg-speedup-thr-exp-1-1m.csv};
  \addplot+[color=black,mark=diamond,style=densely dashdotted] 
  table[x=c1,y=c5,col sep=semicolon] {anc/avg-speedup-thr-exp-1-1m.csv};
\end{loglogaxis}
\end{tikzpicture}
\end{minipage}
\ref{leg:avg_cpu_time_threads_exp_1}
\caption{Benchmark 1 on Intel Xeon: Average CPU time to compute a single time step (left) and average speedup in relation to the fastest serial time (right) for several thread counts and mesh granularities of a single-layered domain using polynomial order 5. Values are shown in Table~\ref{tab:avg_cpu_time_threads_exp_1}. Time, speedup and threads are in logarithmic scale.}
\label{fig:avg_cpu_time_threads_exp_1}
\end{figure*}

We divided the first benchmark into three parts. In the first part, we simulated the propagation of an acoustic wave in a single-layered domain to assess the average CPU time required to compute a single time step of the wave. The medium density, compression modulus and wave speed were constant throughout the domain. We used a GNU/Linux system running CentOS 6.5 64-bit equipped with an Intel Xeon \mbox{E5-2698} v3 CPU, with sixteen physical cores and two hardware threads per core, for 32 threads in total. All cores ran at 2.3 GHz and total cache size was 40 MB. We used six different thread counts between 1 and 32, and three different mesh granularities: 250 thousand, 500 thousand and 1 million elements. Element coloring resulted in 11 colors (first and third meshes) and 12 colors (second mesh). The domain was 2000 units wide and 1000 units deep, resulting in SFC depths of 16 (first mesh) and 18 (second and third meshes). Each mesh, although unstructured, had elements of largely homogeneous edge size. We used a polynomial order of 5, resulting in 21 local nodes per element. The execution results are listed in 
Table~\ref{tab:avg_cpu_time_threads_exp_1} and plotted in Figure~\ref{fig:avg_cpu_time_threads_exp_1}.

\begin{table*}[!htb]
\footnotesize
\begin{tabular}
{ c c c c c | c c c c | c c c c }
\toprule
        & \multicolumn{12}{c}{Average CPU time (s) per time step} \\ \cmidrule(){2-13}
        & \multicolumn{4}{c |}{250k-element domain}
        & \multicolumn{4}{c |}{500k-element domain}
        & \multicolumn{4}{c}{1M-element domain} \\
          \cmidrule(){2-5} \cmidrule(){6-9} \cmidrule(){10-13}
        & \multicolumn{4}{c |}{Reordering strategy}
        & \multicolumn{4}{c |}{Reordering strategy}
        & \multicolumn{4}{c}{Reordering strategy} \\
          \cmidrule(){2-5} \cmidrule(){6-9} \cmidrule(){10-13}
Threads & None & Conn. & Dist. & SFC & None & Conn. & Dist. & SFC & None & Conn. & Dist. & SFC \\
\midrule
1  & 5.292 & 5.786 & 5.773 & 4.884 & 10.977 & 11.813 & 11.782 & 10.016 & 22.847 & 24.141 & 24.101 & 20.415 \\
2  & 4.348 & 4.454 & 4.449 & 4.015 &  8.982 &  9.127 &  9.100 &  8.225 & 18.411 & 18.425 & 18.382 & 16.517 \\
4  & 2.368 & 2.283 & 2.275 & 2.054 &  4.905 &  4.711 &  4.700 &  4.276 & 10.041 &  9.542 &  9.522 &  8.639 \\
8  & 1.307 & 1.176 & 1.185 & 1.066 &  2.704 &  2.474 &  2.468 &  2.235 &  5.519 &  5.005 &  5.006 &  4.527 \\
16 & 0.721 & 0.628 & 0.627 & 0.570 &  1.490 &  1.333 &  1.333 &  1.228 &  3.057 &  2.696 &  2.714 &  2.483 \\
32 & 0.436 & 0.359 & 0.354 & 0.327 &  0.867 &  0.772 &  0.775 &  0.697 &  1.736 &  1.576 &  1.568 &  1.465 \\
\bottomrule
\end{tabular}
\centering
\caption{Benchmark 1 on Intel Xeon: Average CPU time to compute a single time step in a 32-thread system for several thread counts and mesh granularities of a single-layered domain using polynomial order~5.}
\label{tab:avg_cpu_time_threads_exp_1}
\end{table*}
\normalsize

The CPU times in Table \ref{tab:avg_cpu_time_threads_exp_1} and Figure \ref{fig:avg_cpu_time_threads_exp_1} show that the ``SFC'' strategy was always the fastest option. It was most advantageous when compared to the ``None'' strategy, especially when using a larger number of threads. We highlight the fact that the ``Conn.'' and ``Dist.'' strategies were sometimes slower than the ``None'' strategy, as seen in the results for one and two threads. Therefore, when developing reordering strategies, we must consider that performance may actually decline in specific cases. Another observation from Figure~\ref{fig:avg_cpu_time_threads_exp_1} is that the speedup of SFC-based ordering in relation to the fastest serial time for each mesh granularity was always higher than all other strategies.

In the second part of the first benchmark, we ran simulations with several polynomial orders and mesh granularities of the single-layered domain using all available $32$ threads. The results are shown in Table \ref{tab:avg_cpu_time_orders_exp_1} and plotted in Figure \ref{fig:avg_cpu_time_orders_exp_1}. The ``SFC'' strategy retained the best performance, running between 14.8\% (one million elements, order 7) and 25.1\% (250 thousand elements, order 5) faster than the ``None'' strategy. However we must point that, as the mesh granularity increased, the higher data volume per element did not benefit from using SFC-based ordering as much.

\begin{table*}[!htb]
\small
\begin{tabular}
{ c c c c c | c c c c | c c c c }
\toprule
        & \multicolumn{12}{c}{Average CPU time (s) per time step,} \\
        & \multicolumn{12}{c}{percentage of time in relation to ``None'' reordering strategy} \\ \cmidrule(){2-13}
        & \multicolumn{4}{c |}{250k-element domain}
        & \multicolumn{4}{c |}{500k-element domain}
        & \multicolumn{4}{c}{1M-element domain} \\
          \cmidrule(){2-5} \cmidrule(){6-9} \cmidrule(){10-13}
        & \multicolumn{4}{c |}{Reordering strategy}
        & \multicolumn{4}{c |}{Reordering strategy}
        & \multicolumn{4}{c}{Reordering strategy} \\
          \cmidrule(){2-5} \cmidrule(){6-9} \cmidrule(){10-13}
Order & None & Conn. & Dist. & SFC & None & Conn. & Dist. & SFC & None & Conn. & Dist. & SFC \\
\midrule
5 & 0.436, & 0.359, & 0.354, & 0.327, & 0.867, & 0.772, & 0.775, & 0.697, & 1.736, & 1.576, & 1.568, & 1.465, \\
  & -      & 82.4\% & 81.1\% & 74.9\% & -      & 89.1\% & 89.4\% & 80.5\% & -      & 90.8\% & 90.4\% & 84.4\% \\
6 & 0.644, & 0.533, & 0.531, & 0.514, & 1.285, & 1.117, & 1.119, & 1.080, & 2.646, & 2.344, & 2.355, & 2.238, \\
  & -      & 82.8\% & 82.4\% & 79.9\% & -      & 86.9\% & 87.1\% & 84.1\% & -      & 88.6\% & 89.0\% & 84.6\% \\
7 & 0.970, & 0.804, & 0.809, & 0.775, & 1.967, & 1.681, & 1.680, & 1.618, & 4.010, & 3.541, & 3.575, & 3.414, \\
  & -      & 82.9\% & 83.4\% & 79.8\% & -      & 85.4\% & 85.4\% & 82.2\% & -      & 88.3\% & 89.2\% & 85.2\% \\
\bottomrule
\end{tabular}
\centering
\caption{Benchmark 1 on Intel Xeon: Average CPU time to compute a single time step and percentage of time in relation to ``None'' reordering strategy, for several polynomial orders and mesh granularities of a single-layered domain using 32 threads.}
\label{tab:avg_cpu_time_orders_exp_1}
\end{table*}
\normalsize

\begin{figure}[!htb]
\noindent
\centering
\begin{tikzpicture}
\begin{axis}[
  height=8.0cm,
  width=\linewidth,
  grid=major, grid style={dashed,gray!30},
  xlabel=Polynomial order, ylabel=Time (s),
  legend columns=2, legend pos=north west,
  xtick={ 5, 6, 7 },
  log ticks with fixed point,
  ymax=45, ymode=log
  ]
  \addplot+[color=blue,mark=square,style=solid]
  table[x=c1,y=c2,col sep=semicolon] {anc/avg-cpu-time-orders-exp-1-250k.csv};
  \addlegendentry{\scriptsize None (250k)}
  \addplot+[color=red,mark=triangle,style=solid]
  table[x=c1,y=c3,col sep=semicolon] {anc/avg-cpu-time-orders-exp-1-250k.csv};
  \addlegendentry{\scriptsize Conn. (250k)}
  \addplot+[color=purple,mark=o,style=solid]
  table[x=c1,y=c4,col sep=semicolon] {anc/avg-cpu-time-orders-exp-1-250k.csv};
  \addlegendentry{\scriptsize Dist. (250k)}
  \addplot+[color=black,mark=diamond,style=solid]
  table[x=c1,y=c5,col sep=semicolon] {anc/avg-cpu-time-orders-exp-1-250k.csv};
  \addlegendentry{\scriptsize SFC (250k)}

  \addplot+[color=blue,mark=square,style=densely dotted] 
  table[x=c1,y=c2,col sep=semicolon] {anc/avg-cpu-time-orders-exp-1-500k.csv};
  \addlegendentry{\scriptsize None (500k)}
  \addplot+[color=red,mark=triangle,style=densely dotted] 
  table[x=c1,y=c3,col sep=semicolon] {anc/avg-cpu-time-orders-exp-1-500k.csv};
  \addlegendentry{\scriptsize Conn. (500k)}
  \addplot+[color=purple,mark=o,style=densely dotted] 
  table[x=c1,y=c4,col sep=semicolon] {anc/avg-cpu-time-orders-exp-1-500k.csv};
  \addlegendentry{\scriptsize Dist. (500k)}
  \addplot+[color=black,mark=diamond,style=densely dotted] 
  table[x=c1,y=c5,col sep=semicolon] {anc/avg-cpu-time-orders-exp-1-500k.csv};
  \addlegendentry{\scriptsize SFC (500k)}

  \addplot+[color=blue,mark=square,style=densely dashdotted] 
  table[x=c1,y=c2,col sep=semicolon] {anc/avg-cpu-time-orders-exp-1-1m.csv};
  \addlegendentry{\scriptsize None (1M)}
  \addplot+[color=red,mark=triangle,style=densely dashdotted] 
  table[x=c1,y=c3,col sep=semicolon] {anc/avg-cpu-time-orders-exp-1-1m.csv};
  \addlegendentry{\scriptsize Conn. (1M)}
  \addplot+[color=purple,mark=o,style=densely dashdotted] 
  table[x=c1,y=c4,col sep=semicolon] {anc/avg-cpu-time-orders-exp-1-1m.csv};
  \addlegendentry{\scriptsize Dist. (1M)}
  \addplot+[color=black,mark=diamond,style=densely dashdotted] 
  table[x=c1,y=c5,col sep=semicolon] {anc/avg-cpu-time-orders-exp-1-1m.csv};
  \addlegendentry{\scriptsize SFC (1M)}
\end{axis}
\end{tikzpicture}
\caption{Benchmark 1 on Intel Xeon: Average CPU time to compute a single time step for several polynomial orders and mesh granularities of a single-layered domain using 32 threads. Values are shown in Table~\ref{tab:avg_cpu_time_orders_exp_1}. Time is in logarithmic scale.}
\label{fig:avg_cpu_time_orders_exp_1}
\end{figure}

\begin{figure}[!htb]
\noindent
\centering
\begin{tikzpicture}
\begin{axis}[
  height=8.0cm,
  width=0.90 \linewidth,
  grid=major, grid style={dashed,gray!30},
  xlabel=Mesh elements, ylabel=Time (s),
  legend columns=1, 
  legend style={at={(0.02,0.71)},anchor=north west},
  symbolic x coords={ $250k$, $500k$, $1M$ },
  xtick={ $250k$, $500k$, $1M$ }, ymax=7.5,
  axis y line*=left
  ]
  \addplot+[color=blue,mark=square,style=solid]
  table[x=c1,y=c2,col sep=semicolon] {anc/mem-data-exp-1-time.csv};
  \addlegendentry{\scriptsize None (time)}
  \addplot+[color=red,mark=triangle,style=solid]
  table[x=c1,y=c3,col sep=semicolon] {anc/mem-data-exp-1-time.csv};
  \addlegendentry{\scriptsize Conn. (time)}
  \addplot+[color=purple,mark=o,style=solid]
  table[x=c1,y=c4,col sep=semicolon] {anc/mem-data-exp-1-time.csv};
  \addlegendentry{\scriptsize Dist. (time)}
  \addplot+[color=black,mark=diamond,style=solid]
  table[x=c1,y=c5,col sep=semicolon] {anc/mem-data-exp-1-time.csv};
  \addlegendentry{\scriptsize SFC (time)}
\end{axis}
\begin{axis}[
  height=8.0cm,
  width=0.90 \linewidth,
  ylabel=LLC misses (millions),
  legend columns=1, 
  legend style={at={(0.02,0.98)},anchor=north west},
  symbolic x coords={ $250k$, $500k$, $1M$ },
  xtick={ $250k$, $500k$, $1M$ }, 
  scaled y ticks=base 10:-3,
  axis y line*=right,
  axis x line=none
  ]
  \addplot+[color=blue,mark=square,style=densely dashdotted] 
  table[x=c1,y=c2,col sep=semicolon] {anc/mem-data-exp-1-llc.csv};
  \addlegendentry{\scriptsize None (LLC misses)}
  \addplot+[color=red,mark=triangle,style=densely dashdotted] 
  table[x=c1,y=c3,col sep=semicolon] {anc/mem-data-exp-1-llc.csv};
  \addlegendentry{\scriptsize Conn. (LLC misses)}
  \addplot+[color=purple,mark=o,style=densely dashdotted] 
  table[x=c1,y=c4,col sep=semicolon] {anc/mem-data-exp-1-llc.csv};
  \addlegendentry{\scriptsize Dist. (LLC misses)}
  \addplot+[color=black,mark=diamond,style=densely dashdotted] 
  table[x=c1,y=c5,col sep=semicolon] {anc/mem-data-exp-1-llc.csv};
  \addlegendentry{\scriptsize SFC (LLC misses)}
\end{axis}
\end{tikzpicture}
\caption{Benchmark 1 on \mbox{Intel i7}: Average CPU time to compute a single time step and LLC misses in a four-thread system for several mesh granularities of a single-layered domain using polynomial order 5. Values are shown in Table~\ref{tab:mem_data_exp_1}. Mesh elements are in logarithmic scale.}
\label{fig:mem_data_exp_1}
\end{figure}
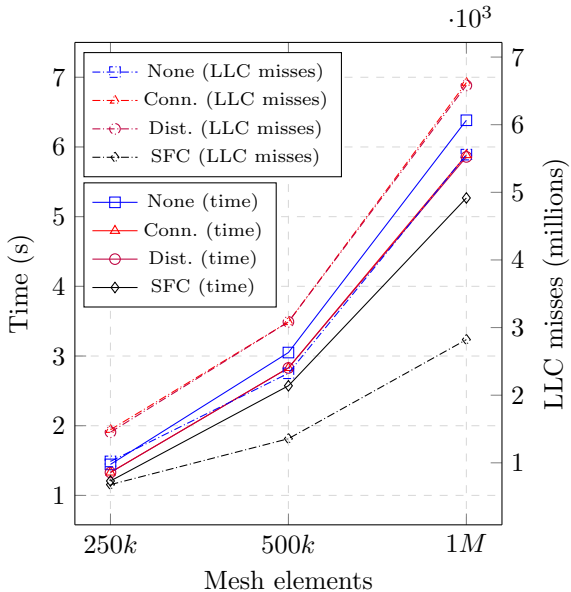

The third and final part of the first benchmark measured the number of last-level cache (LLC) misses and the percentage of stalled slots in the memory pipeline (SSMP) on the single-layered domain. This time we used a GNU/Linux system running Ubuntu 18.04.5 LTS equipped with an Intel \mbox{i7-7500U} CPU, with two physical cores and two hardware threads per core, for four threads in total. All cores ran at 3.5 GHz and total cache size was 4 MB. Polynomial order was fixed at 5. We made measurements using Intel VTune Profiler \parencite*{Vtune20}, and they are shown in Table \ref{tab:mem_data_exp_1} and Figure \ref{fig:mem_data_exp_1}. SFC-based ordering once again yielded the lowest compute time, with the corresponding number of LLC misses and SSMP percentage significantly lower than the other strategies. It also stands out that, under the ``Conn.'' and ``Dist.'' strategies, LLC misses were noticeably higher than those of the ``None'' strategy, but all SSMP percentages were slightly lower and performance was always better.

\begin{table}[!htb]
\small
\begin{tabular}
{ c c c c c }
\toprule
         & \multicolumn{4}{c}{Average CPU time (s),} \\
         & \multicolumn{4}{c}{LLC misses (millions),} \\
         & \multicolumn{4}{c}{SSMP percentage} \\ \cmidrule(){2-5}
         & \multicolumn{4}{c}{Reordering strategy} \\ \cmidrule(){2-5}
Domain   & None & Conn. & Dist. & SFC \\
elements &      &       &       &     \\
\midrule
250k & 1.445,   & 1.330,   & 1.329,   & 1.207,   \\
     & 1021.03, & 1489.78, & 1453.06, & 677.57,  \\
     & 14.9\%   & 14.7\%   & 14.4\%   & 8.3\%    \\
500k & 3.050,   & 2.830,   & 2.829,   & 2.573,   \\
     & 2324.32, & 3083.26, & 3083.98, & 1356.57, \\
     & 15.1\%   & 14.4\%   & 14.3\%   & 7.4\%    \\
1M   & 6.383,   & 5.885,   & 5.857,   & 5.268,   \\
     & 5557.35, & 6622.30, & 6580.54, & 2821.88, \\
     & 16.4\%   & 15.0\%   & 14.8\%   & 7.3\%    \\
\midrule
\multicolumn{5}{l}{LLC = last-level cache} \\
\multicolumn{5}{l}{SSMP = stalled slots in the memory pipeline} \\
\bottomrule
\end{tabular}
\centering
\caption{Benchmark 1 on \mbox{Intel i7}: Average CPU time to compute a single time step, LLC misses and SSMP percentage in a four-thread system for several mesh granularities of a single-layered domain using polynomial order 5.}
\label{tab:mem_data_exp_1}
\end{table}
\normalsize

\subsubsection{Multilayered Domain}

\begin{figure*}[!htb]
\centering
  \begin{minipage}[b]{0.85\linewidth}
    \centering
    \resizebox{1.0\linewidth}{!}{\includegraphics{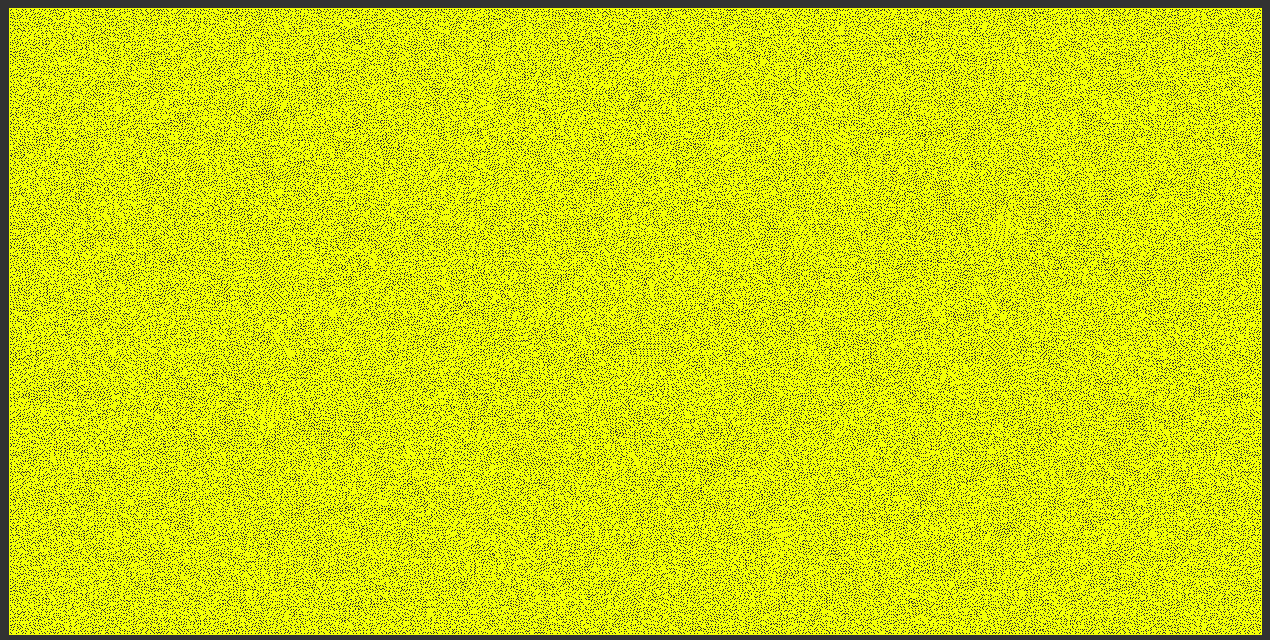}}
  \end{minipage}
  \begin{minipage}[b]{0.85\linewidth}
    \centering
    \resizebox{1.0\linewidth}{!}{\includegraphics{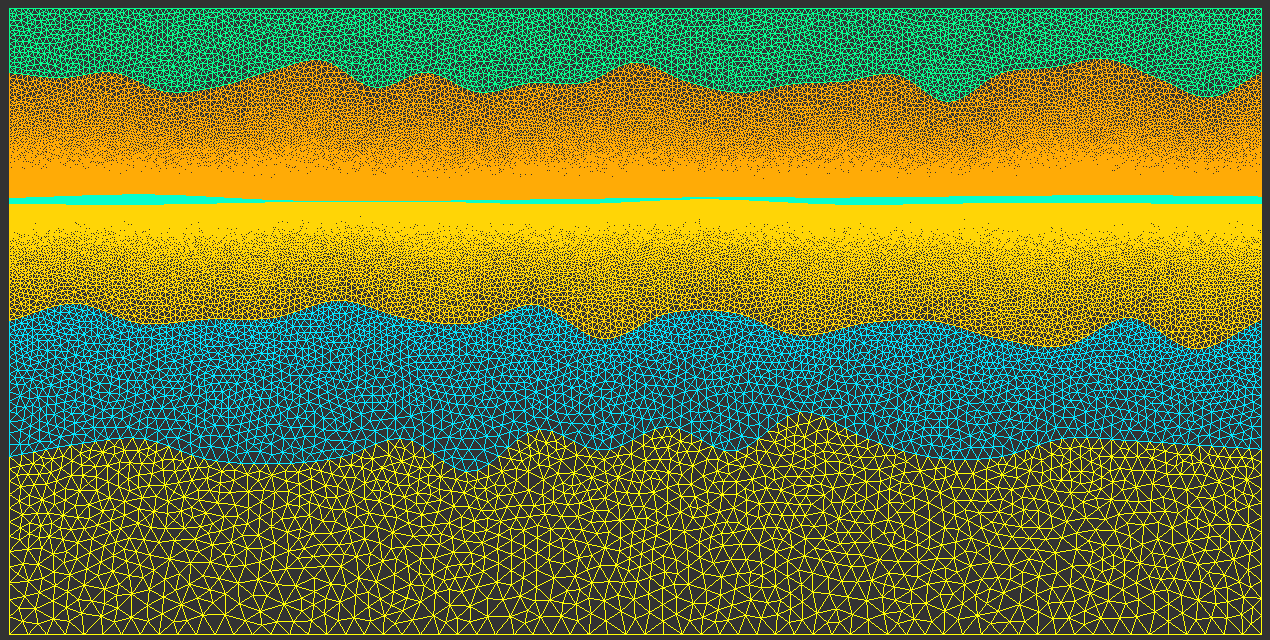}}
  \end{minipage}
  \caption{Meshes with 250 thousand elements used in Benchmark 1 (above) and Benchmark 2 (below). Whereas element sizes in the meshes used in the first benchmark were largely homogeneous, in the second benchmark element sizes had much higher variation.}
  \label{fig:compare-meshes-exp-1-2}
\end{figure*}

In the second benchmark, divided into three parts following those of the first benchmark, we simulated the propagation of an acoustic wave in a multilayered domain with six different layers of varying geometric complexity and properties. Defining the size of an element as the length of its largest edge, the variation in element sizes in the meshes used for the second benchmark was much higher than that of the meshes used in the first benchmark, as shown in Table \ref{tab:mesh_variation} and Figure \ref{fig:compare-meshes-exp-1-2}. As before, the domain was 2000 units wide and 1000 units deep. In the second benchmark, the first mesh had 12 colors and depth 16, and the third and second meshes had 11 colors and depth 18. The purpose of this benchmark was to assess whether larger differences in element sizes would lead to significant leaps between spatially distant elements when traversing them through SFC-based orderings, such that these leaps would translate into worse data locality and execution performance.

\begin{table}[!htb]
\begin{tabular}
{ c c c }
\toprule
         & \multicolumn{2}{c}{Standard deviation} \\
         & \multicolumn{2}{c}{in element size,} \\
         & \multicolumn{2}{c}{ratio between maximum} \\
         & \multicolumn{2}{c}{and minimum element size} \\ \cmidrule(){2-3}
Mesh     & Benchmark & Benchmark \\
Elements & 1         & 2 \\
\midrule
250k & 0.58, & 3.79, \\
     & 2.63  & 62.01 \\
500k & 0.41, & 2.51, \\
     & 2.61  & 55.99 \\
1M   & 0.29, & 1.55, \\
     & 2.81  & 29.91 \\
\bottomrule
\end{tabular}
\centering
\caption{Variation in element sizes in the first and second benchmarks for several mesh granularities.}
\label{tab:mesh_variation}
\end{table}

\begin{figure*}[!htb]
\noindent
\centering
\begin{minipage}[b]{0.49\linewidth}
\begin{tikzpicture}
\begin{loglogaxis}[
  legend style={font=\footnotesize},
  legend pos=north west,
  legend columns=4,
  legend to name=leg:avg_time_per_elem_thr_scatter,
  height=7.0cm,
  width=\linewidth,
  grid=major, grid style={dashed,gray!30},
  ylabel=Time per element ($\mu$s) -- Benchmark 1,
  xlabel=Time per element ($\mu$s) -- Benchmark 2,
  log ticks with fixed point,
  scaled x ticks=base 10:-6,
  scaled y ticks=base 10:-6
  ]
  \addplot+[
    scatter/classes={
      none250={color=blue,mark=square,style=solid},
      conn250={color=red,mark=triangle,style=solid},
      dist250={color=purple,mark=o,style=solid},
      sfc250={color=black,mark=diamond,style=solid},
      none500={color=blue,mark=square,style=densely dotted},
      conn500={color=red,mark=triangle,style=densely dotted},
      dist500={color=purple,mark=o,style=densely dotted},
      sfc500={color=black,mark=diamond,style=densely dotted},
      none1m={color=blue,mark=square,style=densely dashdotted},
      conn1m={color=red,mark=triangle,style=densely dashdotted},
      dist1m={color=purple,mark=o,style=densely dashdotted},
      sfc1m={color=black,mark=diamond,style=densely dashdotted}
    },
    scatter, only marks,
    scatter src=explicit symbolic
  ]
  table[y=ex1, x=ex2, meta=label] {anc/avg-time-per-elem-thr-scatter.dat};
  \addplot+[color=black,mark=none] coordinates
    {(0.8, 0.8) (40.0, 40.0)};
  \legend{None (250k), Conn. (250k) , Dist. (250k), SFC (250k),
    None (500k), Conn. (500k) , Dist. (500k), SFC (500k),
    None (1M), Conn. (1M) , Dist. (1M), SFC (1M)}
\end{loglogaxis}
\end{tikzpicture}
\end{minipage}
\begin{minipage}[b]{0.49\linewidth}
\centering
\begin{tikzpicture}
\begin{loglogaxis}[
  legend style={font=\scriptsize},
  legend pos=north west,
  legend columns=2,
  height=7.0cm,
  width=\linewidth,
  grid=major, grid style={dashed,gray!30},
  ylabel=Speedup -- Benchmark 1,
  xlabel=Speedup -- Benchmark 2,
  log ticks with fixed point
  ]
  \addplot+[
    scatter/classes={
      none250={color=blue,mark=square,style=solid},
      conn250={color=red,mark=triangle,style=solid},
      dist250={color=purple,mark=o,style=solid},
      sfc250={color=black,mark=diamond,style=solid},
      none500={color=blue,mark=square,style=densely dotted},
      conn500={color=red,mark=triangle,style=densely dotted},
      dist500={color=purple,mark=o,style=densely dotted},
      sfc500={color=black,mark=diamond,style=densely dotted},
      none1m={color=blue,mark=square,style=densely dashdotted},
      conn1m={color=red,mark=triangle,style=densely dashdotted},
      dist1m={color=purple,mark=o,style=densely dashdotted},
      sfc1m={color=black,mark=diamond,style=densely dashdotted}
    },
    scatter, only marks,
    scatter src=explicit symbolic
  ]
  table[y=ex1, x=ex2, meta=label] {anc/avg-speedup-thr-scatter.dat};
  \addplot+[color=black,mark=none] coordinates
    {(0.8, 0.8) (20.0, 20.0)};
\end{loglogaxis}
\end{tikzpicture}
\end{minipage}
\ref{leg:avg_time_per_elem_thr_scatter}
\caption{Execution time per element for Benchmark 1 versus Benchmark 2 (left) and average speedup in relation to the best serial time for Benchmark 1 versus Benchmark 2 (right), on Intel Xeon, for several thread counts and mesh granularities using polynomial order 5. Overall, the times per element are higher in Benchmark 2 and the speedups of SFC-based ordering are higher in Benchmark 1. The diagonal lines show where the measurements of Benchmarks 1 and 2 would be located if both benchmarks had identical performance. Time per element and speedup are in logarithmic scale.}
\label{fig:avg_time_per_elem_thr_scatter}
\end{figure*}
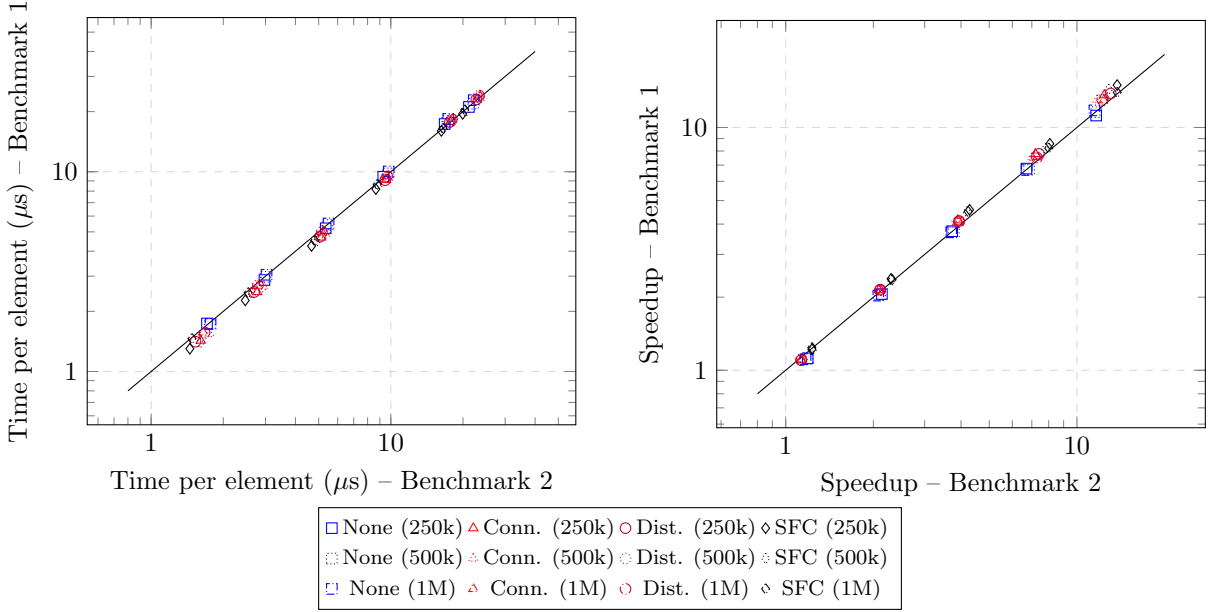

The CPU times for the first set of benchmarks, testing different mesh granularities and reordering strategies for different thread counts using polynomial order 5, are listed in Table \ref{tab:avg_cpu_time_threads_exp_2}. They should be compared to the corresponding Table \ref{tab:avg_cpu_time_threads_exp_1} of the first benchmark. Figure \ref{fig:avg_cpu_time_threads_exp_2} shows the times and speedups of Table \ref{tab:avg_cpu_time_threads_exp_2}, whereas Figure \ref{fig:avg_time_per_elem_thr_scatter} compares the times per element and speedups between the two benchmarks.
SFC-based orderings maintained the clear performance advantage seen in the first benchmark, but to a smaller extent. This is caused by the greater variation in element sizes in the meshes of the second benchmark: upon following the steps of the Hilbert curve to iterate between two elements, if these elements are spatially distant from each other, the number of elements in the vicinity of the second element sharing nodes with the first element is likely to be lower. Therefore, significant chunks of the node data loaded in cache lines for the previous element cannot be used by the elements following it, causing cache misses and worse memory efficiency as new nodes need to be read into cache to process the next element.

\begin{table*}[!htb]
\footnotesize
\begin{tabular}
{ c c c c c | c c c c | c c c c }
\toprule
        & \multicolumn{12}{c}{Average CPU time (s) per time step} \\ \cmidrule(){2-13}
        & \multicolumn{4}{c |}{250k-element domain}
        & \multicolumn{4}{c |}{500k-element domain}
        & \multicolumn{4}{c}{1M-element domain} \\
          \cmidrule(){2-5} \cmidrule(){6-9} \cmidrule(){10-13}
        & \multicolumn{4}{c |}{Reordering strategy}
        & \multicolumn{4}{c |}{Reordering strategy}
        & \multicolumn{4}{c}{Reordering strategy} \\
          \cmidrule(){2-5} \cmidrule(){6-9} \cmidrule(){10-13}
Threads & None & Conn. & Dist. & SFC & None & Conn. & Dist. & SFC & None & Conn. & Dist. & SFC \\
\midrule
1  & 5.307 & 5.721 & 5.696 & 4.998 & 11.025 & 11.769 & 11.719 & 10.282 & 22.205 & 23.433 & 23.334 & 20.361 \\
2  & 4.203 & 4.439 & 4.417 & 4.070 &  8.608 &  8.992 &  8.992 &  8.311 & 17.369 & 18.029 & 17.995 & 16.517 \\
4  & 2.331 & 2.390 & 2.373 & 2.170 &  4.867 &  4.839 &  4.865 &  4.449 &  9.802 &  9.622 &  9.654 &  8.770 \\
8  & 1.341 & 1.283 & 1.268 & 1.169 &  2.723 &  2.647 &  2.651 &  2.451 &  5.505 &  5.159 &  5.166 &  4.814 \\
16 & 0.745 & 0.694 & 0.675 & 0.620 &  1.502 &  1.463 &  1.439 &  1.320 &  3.030 &  2.779 &  2.795 &  2.544 \\
32 & 0.429 & 0.403 & 0.383 & 0.364 &  0.882 &  0.876 &  0.836 &  0.800 &  1.774 &  1.646 &  1.667 &  1.481 \\
\bottomrule
\end{tabular}
\centering
\caption{Benchmark 2 on Intel Xeon: Average CPU time to compute a single time step in a 32-thread system for several thread counts and mesh granularities of a multilayered domain using polynomial order 5.}
\label{tab:avg_cpu_time_threads_exp_2}
\end{table*}
\normalsize

\begin{figure*}[!htb]
\noindent
\centering
\begin{minipage}[b]{0.49\linewidth}
\begin{tikzpicture}
\begin{loglogaxis}[
  height=7.0cm,
  width=\linewidth,
  grid=major, grid style={dashed,gray!30},
  xlabel=Threads, ylabel=CPU Time (s),
  xtick={ 1, 2, 4, 8, 16, 32 },
  log ticks with fixed point,
  legend columns=4,
  legend to name=leg:avg_cpu_time_threads_exp_2
  ]
  \addplot+[color=blue,mark=square,style=solid]
  table[x=c1,y=c2,col sep=semicolon] {anc/avg-cpu-time-thr-exp-2-250k.csv};
  \addlegendentry{\footnotesize  None (250k)}
  \addplot+[color=red,mark=triangle,style=solid]
  table[x=c1,y=c3,col sep=semicolon] {anc/avg-cpu-time-thr-exp-2-250k.csv};
  \addlegendentry{\footnotesize  Conn. (250k)}
  \addplot+[color=purple,mark=o,style=solid]
  table[x=c1,y=c4,col sep=semicolon] {anc/avg-cpu-time-thr-exp-2-250k.csv};
  \addlegendentry{\footnotesize  Dist. (250k)}
  \addplot+[color=black,mark=diamond,style=solid]
  table[x=c1,y=c5,col sep=semicolon] {anc/avg-cpu-time-thr-exp-2-250k.csv};
  \addlegendentry{\footnotesize  SFC (250k)}

  \addplot+[color=blue,mark=square,style=densely dotted] 
  table[x=c1,y=c2,col sep=semicolon] {anc/avg-cpu-time-thr-exp-2-500k.csv};
  \addlegendentry{\footnotesize None (500k)}
  \addplot+[color=red,mark=triangle,style=densely dotted] 
  table[x=c1,y=c3,col sep=semicolon] {anc/avg-cpu-time-thr-exp-2-500k.csv};
  \addlegendentry{\footnotesize Conn. (500k)}
  \addplot+[color=purple,mark=o,style=densely dotted] 
  table[x=c1,y=c4,col sep=semicolon] {anc/avg-cpu-time-thr-exp-2-500k.csv};
  \addlegendentry{\footnotesize Dist. (500k)}
  \addplot+[color=black,mark=diamond,style=densely dotted] 
  table[x=c1,y=c5,col sep=semicolon] {anc/avg-cpu-time-thr-exp-2-500k.csv};
  \addlegendentry{\footnotesize SFC (500k)}

  \addplot+[color=blue,mark=square,style=densely dashdotted] 
  table[x=c1,y=c2,col sep=semicolon] {anc/avg-cpu-time-thr-exp-2-1m.csv};
  \addlegendentry{\footnotesize None (1M)}
  \addplot+[color=red,mark=triangle,style=densely dashdotted] 
  table[x=c1,y=c3,col sep=semicolon] {anc/avg-cpu-time-thr-exp-2-1m.csv};
  \addlegendentry{\footnotesize Conn. (1M)}
  \addplot+[color=purple,mark=o,style=densely dashdotted] 
  table[x=c1,y=c4,col sep=semicolon] {anc/avg-cpu-time-thr-exp-2-1m.csv};
  \addlegendentry{\footnotesize Dist. (1M)}
  \addplot+[color=black,mark=diamond,style=densely dashdotted] 
  table[x=c1,y=c5,col sep=semicolon] {anc/avg-cpu-time-thr-exp-2-1m.csv};
  \addlegendentry{\footnotesize SFC (1M)}
\end{loglogaxis}
\end{tikzpicture}
\end{minipage}
\begin{minipage}[b]{0.49\linewidth}
\centering
\begin{tikzpicture}
\begin{loglogaxis}[
  height=7.0cm,
  width=\linewidth,
  grid=major, grid style={dashed,gray!30},
  xlabel=Threads, ylabel=Speedup,
  legend columns=2, legend pos=north west,
  xtick={ 1, 2, 4, 8, 16, 32 },
  log ticks with fixed point
  ]
  \addplot+[color=blue,mark=square,style=solid]
  table[x=c1,y=c2,col sep=semicolon] {anc/avg-speedup-thr-exp-2-250k.csv};
  \addplot+[color=red,mark=triangle,style=solid]
  table[x=c1,y=c3,col sep=semicolon] {anc/avg-speedup-thr-exp-2-250k.csv};
  \addplot+[color=purple,mark=o,style=solid]
  table[x=c1,y=c4,col sep=semicolon] {anc/avg-speedup-thr-exp-2-250k.csv};
  \addplot+[color=black,mark=diamond,style=solid]
  table[x=c1,y=c5,col sep=semicolon] {anc/avg-speedup-thr-exp-2-250k.csv};

  \addplot+[color=blue,mark=square,style=densely dotted] 
  table[x=c1,y=c2,col sep=semicolon] {anc/avg-speedup-thr-exp-2-500k.csv};
  \addplot+[color=red,mark=triangle,style=densely dotted] 
  table[x=c1,y=c3,col sep=semicolon] {anc/avg-speedup-thr-exp-2-500k.csv};
  \addplot+[color=purple,mark=o,style=densely dotted] 
  table[x=c1,y=c4,col sep=semicolon] {anc/avg-speedup-thr-exp-2-500k.csv};
  \addplot+[color=black,mark=diamond,style=densely dotted] 
  table[x=c1,y=c5,col sep=semicolon] {anc/avg-speedup-thr-exp-2-500k.csv};

  \addplot+[color=blue,mark=square,style=densely dashdotted] 
  table[x=c1,y=c2,col sep=semicolon] {anc/avg-speedup-thr-exp-2-1m.csv};
  \addplot+[color=red,mark=triangle,style=densely dashdotted] 
  table[x=c1,y=c3,col sep=semicolon] {anc/avg-speedup-thr-exp-2-1m.csv};
  \addplot+[color=purple,mark=o,style=densely dashdotted] 
  table[x=c1,y=c4,col sep=semicolon] {anc/avg-speedup-thr-exp-2-1m.csv};
  \addplot+[color=black,mark=diamond,style=densely dashdotted] 
  table[x=c1,y=c5,col sep=semicolon] {anc/avg-speedup-thr-exp-2-1m.csv};
\end{loglogaxis}
\end{tikzpicture}
\end{minipage}
\ref{leg:avg_cpu_time_threads_exp_2}
\caption{Benchmark 2 on Intel Xeon: Average CPU time to compute a single time step (left) and average speedup in relation to the fastest serial time (right) for several thread counts and mesh granularities of a multilayered domain using polynomial order 5. Values are shown in Table~\ref{tab:avg_cpu_time_threads_exp_2}. Time, speedup and threads are in logarithmic scale.}
\label{fig:avg_cpu_time_threads_exp_2}
\end{figure*}
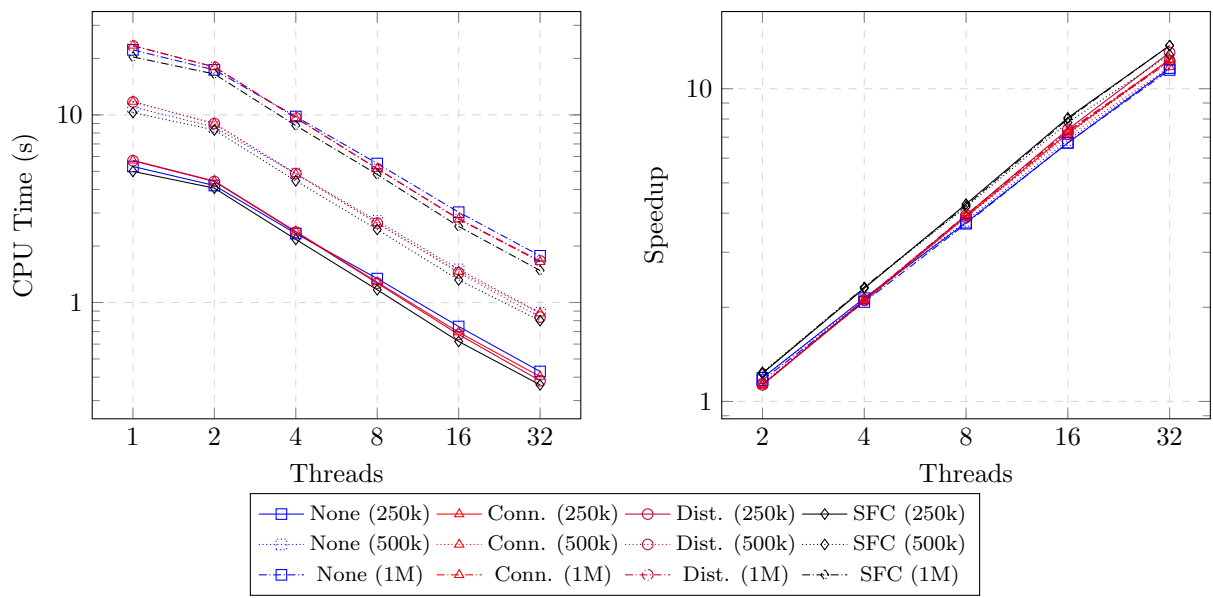

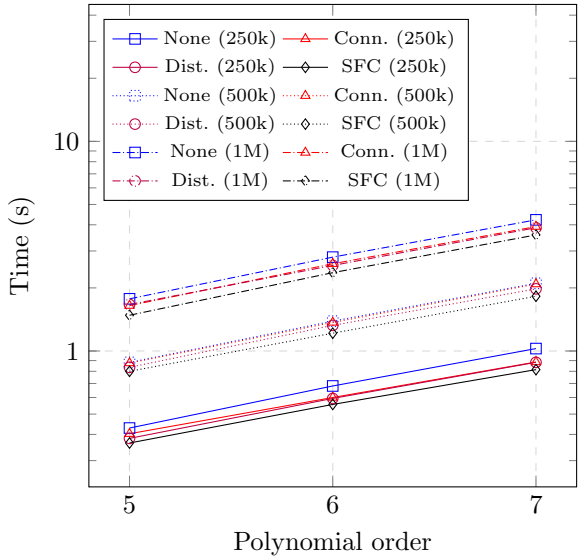
\begin{figure}[!htb]
\noindent
\centering
\begin{tikzpicture}
\begin{axis}[
  height=8.0cm,
  width=\linewidth,
  grid=major, grid style={dashed,gray!30},
  xlabel=Polynomial order, ylabel=Time (s),
  legend columns=2, legend pos=north west,
  xtick={ 5, 6, 7 },
  log ticks with fixed point,
  ymax=45, ymode=log
  ]
  \addplot+[color=blue,mark=square,style=solid]
  table[x=c1,y=c2,col sep=semicolon] {anc/avg-cpu-time-orders-exp-2-250k.csv};
  \addlegendentry{\scriptsize None (250k)}
  \addplot+[color=red,mark=triangle,style=solid]
  table[x=c1,y=c3,col sep=semicolon] {anc/avg-cpu-time-orders-exp-2-250k.csv};
  \addlegendentry{\scriptsize Conn. (250k)}
  \addplot+[color=purple,mark=o,style=solid]
  table[x=c1,y=c4,col sep=semicolon] {anc/avg-cpu-time-orders-exp-2-250k.csv};
  \addlegendentry{\scriptsize Dist. (250k)}
  \addplot+[color=black,mark=diamond,style=solid]
  table[x=c1,y=c5,col sep=semicolon] {anc/avg-cpu-time-orders-exp-2-250k.csv};
  \addlegendentry{\scriptsize SFC (250k)}

  \addplot+[color=blue,mark=square,style=densely dotted] 
  table[x=c1,y=c2,col sep=semicolon] {anc/avg-cpu-time-orders-exp-2-500k.csv};
  \addlegendentry{\scriptsize None (500k)}
  \addplot+[color=red,mark=triangle,style=densely dotted] 
  table[x=c1,y=c3,col sep=semicolon] {anc/avg-cpu-time-orders-exp-2-500k.csv};
  \addlegendentry{\scriptsize Conn. (500k)}
  \addplot+[color=purple,mark=o,style=densely dotted] 
  table[x=c1,y=c4,col sep=semicolon] {anc/avg-cpu-time-orders-exp-2-500k.csv};
  \addlegendentry{\scriptsize Dist. (500k)}
  \addplot+[color=black,mark=diamond,style=densely dotted] 
  table[x=c1,y=c5,col sep=semicolon] {anc/avg-cpu-time-orders-exp-2-500k.csv};
  \addlegendentry{\scriptsize SFC (500k)}

  \addplot+[color=blue,mark=square,style=densely dashdotted] 
  table[x=c1,y=c2,col sep=semicolon] {anc/avg-cpu-time-orders-exp-2-1m.csv};
  \addlegendentry{\scriptsize None (1M)}
  \addplot+[color=red,mark=triangle,style=densely dashdotted] 
  table[x=c1,y=c3,col sep=semicolon] {anc/avg-cpu-time-orders-exp-2-1m.csv};
  \addlegendentry{\scriptsize Conn. (1M)}
  \addplot+[color=purple,mark=o,style=densely dashdotted] 
  table[x=c1,y=c4,col sep=semicolon] {anc/avg-cpu-time-orders-exp-2-1m.csv};
  \addlegendentry{\scriptsize Dist. (1M)}
  \addplot+[color=black,mark=diamond,style=densely dashdotted] 
  table[x=c1,y=c5,col sep=semicolon] {anc/avg-cpu-time-orders-exp-2-1m.csv};
  \addlegendentry{\scriptsize SFC (1M)}
\end{axis}
\end{tikzpicture}
\caption{Benchmark 2 on Intel Xeon: Average CPU time to compute a single time step for several polynomial orders and mesh granularities of a multilayered domain using 32 threads. Values are shown in Table~\ref{tab:avg_cpu_time_orders_exp_2}. Time is in logarithmic scale.}
\label{fig:avg_cpu_time_orders_exp_2}
\end{figure}

The CPU timings for different polynomial orders and mesh granularities for the second part of the second benchmark can be seen in Table \ref{tab:avg_cpu_time_orders_exp_2} and Figure \ref{fig:avg_cpu_time_orders_exp_2}. Once again, the SFC-based ordering was the fastest strategy, running between 9.3\% (500 thousand elements, order 5) and 20.6\% (250 thousand elements, order 7) faster than the ``None'' strategy. When comparing Tables \ref{tab:avg_cpu_time_orders_exp_2} and \ref{tab:avg_cpu_time_orders_exp_1}, the overall observation is a moderate increase in nearly all compute times. This further emphasizes that the greater variation in element sizes was harmful for all strategies in relation to the first benchmark.
Tables \ref{tab:avg_cpu_time_orders_exp_2} and \ref{tab:avg_cpu_time_orders_exp_1} also show that the computation work per element is considerably higher for higher polynomial orders. Our tests gave no indication that this higher computational work per element would reduce the dependence of the computational performance on the element ordering. We can most likely attribute this to the fact that, for higher-order elements, the data volume for gather and scatter operations is also higher.

\begin{table*}[!htb]
\small
\begin{tabular}
{ c c c c c | c c c c | c c c c }
\toprule
        & \multicolumn{12}{c}{Average CPU time (s) per time step,} \\
        & \multicolumn{12}{c}{percentage of time in relation to ``None'' reordering strategy} \\ \cmidrule(){2-13}
        & \multicolumn{4}{c |}{250k-element domain}
        & \multicolumn{4}{c |}{500k-element domain}
        & \multicolumn{4}{c}{1M-element domain} \\
          \cmidrule(){2-5} \cmidrule(){6-9} \cmidrule(){10-13}
        & \multicolumn{4}{c |}{Reordering strategy}
        & \multicolumn{4}{c |}{Reordering strategy}
        & \multicolumn{4}{c}{Reordering strategy} \\
          \cmidrule(){2-5} \cmidrule(){6-9} \cmidrule(){10-13}
Order & None & Conn. & Dist. & SFC & None & Conn. & Dist. & SFC & None & Conn. & Dist. & SFC \\
\midrule
5 & 0.429, & 0.403, & 0.383, & 0.364,  & 0.882, & 0.876, & 0.836, & 0.800, & 1.774, & 1.646, & 1.667, & 1.481, \\
  & -      & 93.8\% & 89.1\% & 84.7\%  & -      & 99.3\% & 94.7\% & 90.7\% & -      & 92.8\% & 94.0\% & 83.5\% \\
6 & 0.681, & 0.600, & 0.594, & 0.556,  & 1.388, & 1.368, & 1.325, & 1.216, & 2.804, & 2.607, & 2.559, & 2.364, \\
  & -      & 88.2\% & 87.2\% & 81.7\%  & -      & 98.5\% & 95.4\% & 87.6\% & -      & 92.9\% & 91.3\% & 84.3\% \\
7 & 1.027, & 0.886, & 0.883, & 0.815,  & 2.103, & 2.085, & 1.979, & 1.825, & 4.223, & 3.921, & 3.873, & 3.579, \\
  & -      & 86.3\% & 86.0\% & 79.4\%  & -      & 99.1\% & 94.1\% & 86.8\% & -      & 92.9\% & 91.7\% & 84.8\% \\
\bottomrule
\end{tabular}
\centering
\caption{Benchmark 2 on Intel Xeon: Average CPU time to compute a single time step and percentage of time in relation to ``None'' reordering strategy, for several polynomial orders and mesh granularities of a multilayered domain using 32 threads.}
\label{tab:avg_cpu_time_orders_exp_2}
\end{table*}
\normalsize

\begin{figure}[!htb]
\noindent
\centering
\begin{tikzpicture}
\begin{axis}[
  height=8.0cm,
  width=0.90 \linewidth,
  grid=major, grid style={dashed,gray!30},
  xlabel=Mesh elements, ylabel=Time (s),
  legend columns=1, 
  legend style={at={(0.02,0.71)},anchor=north west},
  symbolic x coords={ $250k$, $500k$, $1M$ },
  xtick={ $250k$, $500k$, $1M$ }, ymax=7.5,
  axis y line*=left
  ]
  \addplot+[color=blue,mark=square,style=solid]
  table[x=c1,y=c2,col sep=semicolon] {anc/mem-data-exp-2-time.csv};
  \addlegendentry{\scriptsize None (time)}
  \addplot+[color=red,mark=triangle,style=solid]
  table[x=c1,y=c3,col sep=semicolon] {anc/mem-data-exp-2-time.csv};
  \addlegendentry{\scriptsize Conn. (time)}
  \addplot+[color=purple,mark=o,style=solid]
  table[x=c1,y=c4,col sep=semicolon] {anc/mem-data-exp-2-time.csv};
  \addlegendentry{\scriptsize Dist. (time)}
  \addplot+[color=black,mark=diamond,style=solid]
  table[x=c1,y=c5,col sep=semicolon] {anc/mem-data-exp-2-time.csv};
  \addlegendentry{\scriptsize SFC (time)}
\end{axis}
\begin{axis}[
  height=8.0cm,
  width=0.90 \linewidth,
  ylabel=LLC misses (millions),
  legend columns=1, 
  legend style={at={(0.02,0.98)},anchor=north west},
  symbolic x coords={ $250k$, $500k$, $1M$ },
  xtick={ $250k$, $500k$, $1M$ }, 
  scaled y ticks=base 10:-3,
  axis y line*=right,
  axis x line=none
  ]
  \addplot+[color=blue,mark=square,style=densely dashdotted] 
  table[x=c1,y=c2,col sep=semicolon] {anc/mem-data-exp-2-llc.csv};
  \addlegendentry{\scriptsize None (LLC misses)}
  \addplot+[color=red,mark=triangle,style=densely dashdotted] 
  table[x=c1,y=c3,col sep=semicolon] {anc/mem-data-exp-2-llc.csv};
  \addlegendentry{\scriptsize Conn. (LLC misses)}
  \addplot+[color=purple,mark=o,style=densely dashdotted] 
  table[x=c1,y=c4,col sep=semicolon] {anc/mem-data-exp-2-llc.csv};
  \addlegendentry{\scriptsize Dist. (LLC misses)}
  \addplot+[color=black,mark=diamond,style=densely dashdotted] 
  table[x=c1,y=c5,col sep=semicolon] {anc/mem-data-exp-2-llc.csv};
  \addlegendentry{\scriptsize SFC (LLC misses)}
\end{axis}
\end{tikzpicture}
\caption{Benchmark 2 on \mbox{Intel i7}: Average CPU time to compute a single time step and LLC misses in a four-thread system for several mesh granularities of a multilayered domain using polynomial order 5. Values are shown in Table~\ref{tab:mem_data_exp_2}. Mesh elements are in logarithmic scale.}
\label{fig:mem_data_exp_2}
\end{figure}
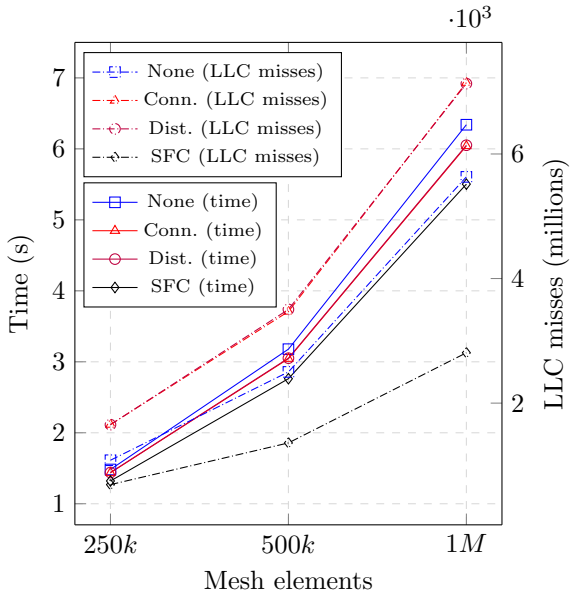

The final part of the second benchmark, which measured average CPU times, LLC misses and SSMP percentages in an \mbox{Intel i7} four-thread system, produced the results seen in Table \ref{tab:mem_data_exp_2} and Figure \ref{fig:mem_data_exp_2}.
The ``SFC'' strategy once again performed the best in all metrics. When comparing Tables \ref{tab:mem_data_exp_2} and \ref{tab:mem_data_exp_1}, we find the execution times of Benchmark 2 to be slightly slower.

\begin{table}[!htb]
\small
\begin{tabular}
{ c c c c c }
\toprule
         & \multicolumn{4}{c}{Average CPU time (s),} \\
         & \multicolumn{4}{c}{LLC misses (millions),} \\
         & \multicolumn{4}{c}{SSMP percentage} \\ \cmidrule(){2-5}
         & \multicolumn{4}{c}{Reordering strategy} \\ \cmidrule(){2-5}
Domain   & None & Conn. & Dist. & SFC \\
elements &      &       &       &     \\
\midrule
250k & 1.483,   & 1.437,   & 1.440,   & 1.321,   \\
     & 1082.96, & 1655.40, & 1653.96, & 691.25,   \\
     & 15.2\%   & 15.8\%   & 14.9\%   & 8.2\%   \\
500k & 3.178,   & 3.044,   & 3.051,   & 2.762,   \\
     & 2500.74, & 3470.64, & 3504.49, & 1360.90,  \\
     & 15.1\%   & 15.5\%   & 14.5\%   & 7.5\%   \\
1M   & 6.340,   & 6.041,   & 6.051,   & 5.503,   \\
     & 5632.23, & 7145.78, & 7129.94, & 2808.20, \\
     & 15.3\%   & 15.1\%   & 15.5\%   & 7.4\%   \\
\midrule
\multicolumn{5}{l}{LLC = last-level cache} \\
\multicolumn{5}{l}{SSMP = stalled slots in the memory pipeline} \\
\bottomrule
\end{tabular}
\centering
\caption{Benchmark 2 on \mbox{Intel i7}: Average CPU time to compute a single time step, LLC misses and SSMP percentage in a four-thread system for several mesh granularities of a multilayered domain using polynomial order 5.}
\label{tab:mem_data_exp_2}
\end{table}
\normalsize

\subsection{Realistic Case Simulation}
\label{sec:realistic_case}

\begin{figure*}[!htb]
  \centering
  \resizebox{.90\linewidth}{!}{\includegraphics{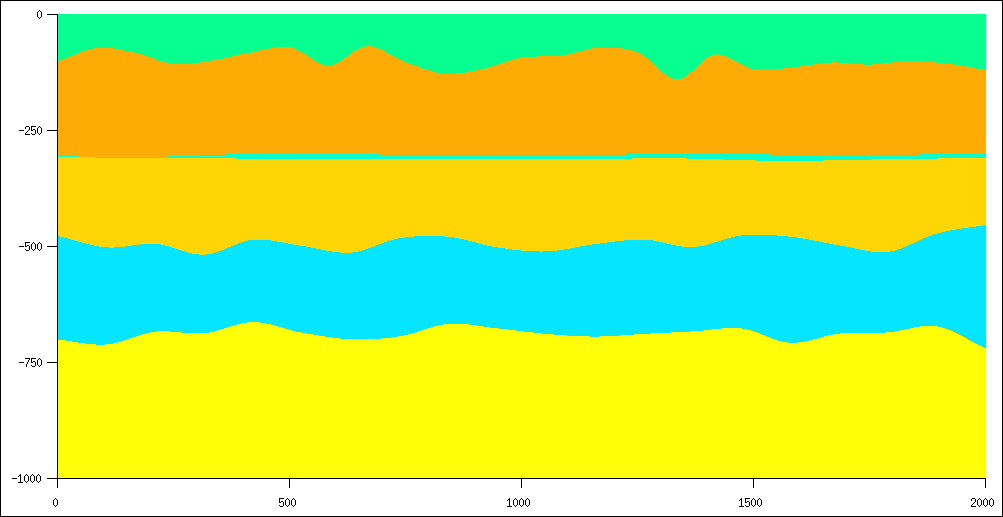}}
  \caption{A multilayered domain. The layer labeling starts from the top with several wave velocities shown in Table~\ref{tab:layer_velocities}. The wave velocities at the deeper layers are higher than those at the shallower layers.}
  \label{fig:complex-domain}
\end{figure*}

Finally we demonstrate the application of the discussed SEM to 2D problems such as synthetic seismic data analysis \parencite{Yilmaz2001} or microseismic data migration \parencite{Trojanowski2017}. We simulated the propagation of a point source with Ricker wavelet profile \parencite{Ricker1944} in a multilayered domain, 2000 units wide and 1000 units deep (see Figure~\ref{fig:complex-domain}), using a one-million element mesh. Each of the six layers of this domain had a different medium density $\rho$ and compression modulus $K$, and therefore a different wave velocity $v_p = \sqrt{K / \rho}$. The wave velocities at the deeper layers were higher than those at the shallower layers (see Table \ref{tab:layer_velocities}). Element size in each layer was chosen proportional to the wave speed of this layer, resulting in a (mostly) constant time step size across the mesh to meet the CFL condition (Courant et al., \cite*{Cfl1967}) for stability of the Heun scheme. The simulation required 197 hours of computing time of 32 threads (sixteen physical cores and two hardware threads per core), with a time step of 1.2 microseconds.

\begin{table}[!htb]
\begin{tabular}
{ c c }
\toprule
Layer & Wave velocity \\
      & (units/s) \\
\midrule
1 & 1500.0 \\
2 & 2500.0 \\
3 & 2000.0 \\
4 & 2500.0 \\
5 & 3000.0 \\
6 & 4000.0 \\
\bottomrule
\end{tabular}
\centering
\caption{Wave velocities at each layer of the multilayered domain. Geometry shown in Figure~\ref{fig:complex-domain}. Labeling of layers starts from the top.}
\label{tab:layer_velocities}
\end{table}

\begin{figure*}[!htb]
\centering
  \begin{minipage}[b]{0.49\linewidth}
    \centering
    \resizebox{1.0\linewidth}{!}{\includegraphics{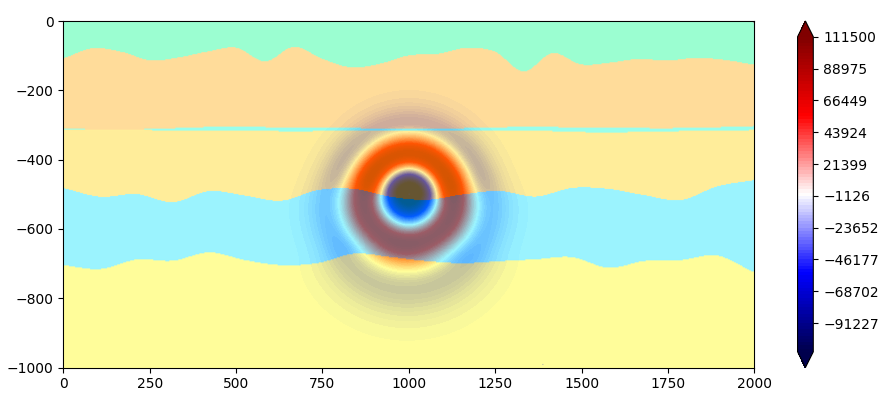}}
  \end{minipage}
  \begin{minipage}[b]{0.49\linewidth}
    \centering
    \resizebox{1.0\linewidth}{!}{\includegraphics{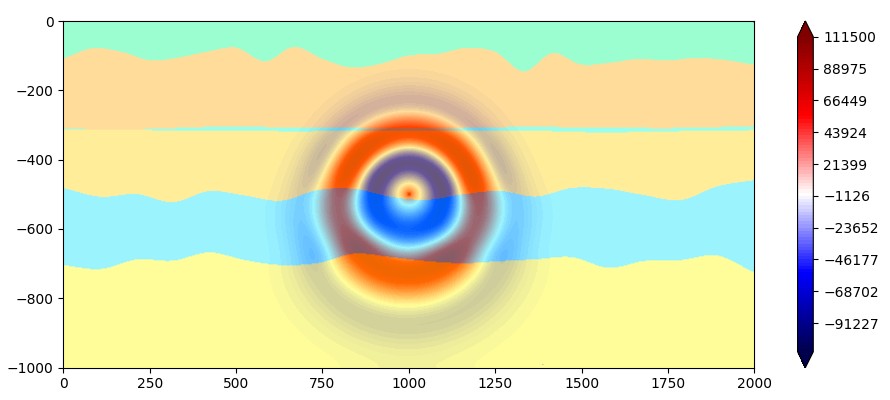}}
  \end{minipage}
  \begin{minipage}[b]{0.49\linewidth}
    \centering
    \resizebox{1.0\linewidth}{!}{\includegraphics{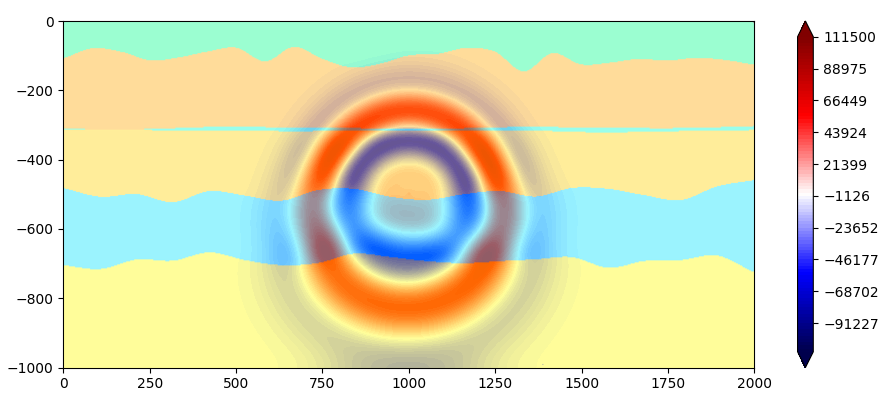}}
  \end{minipage}
  \begin{minipage}[b]{0.49\linewidth}
    \centering
    \resizebox{1.0\linewidth}{!}{\includegraphics{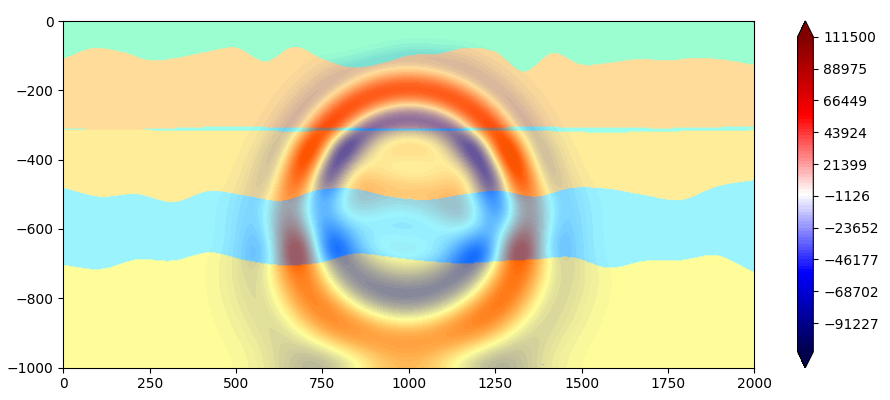}}
  \end{minipage}
  \begin{minipage}[b]{0.49\linewidth}
    \centering
    \resizebox{1.0\linewidth}{!}{\includegraphics{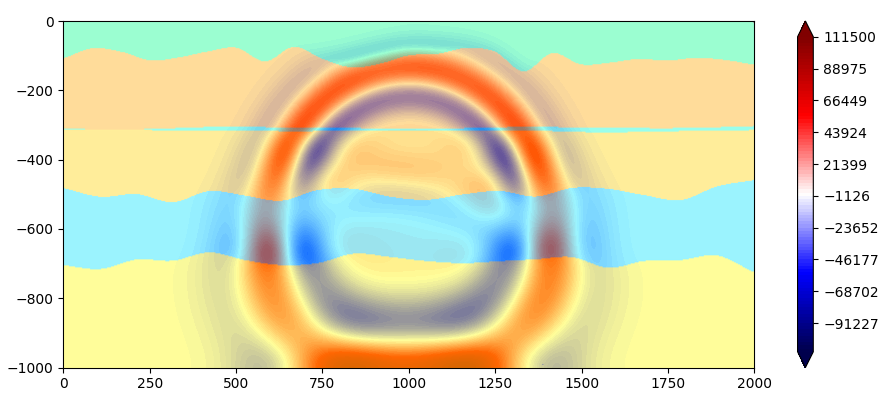}}
  \end{minipage}
  \begin{minipage}[b]{0.49\linewidth}
    \centering
    \resizebox{1.0\linewidth}{!}{\includegraphics{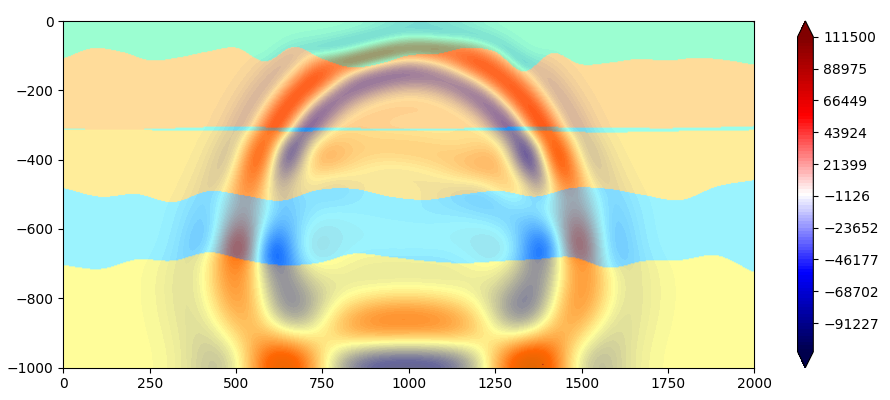}}
  \end{minipage}
  \caption{Ricker wavelet propagation in a multilayered domain at 0.275, 0.300, 0.325, 0.350, 0.375 and 0.400s. The propagation of the wave is noticeably faster at the deeper layers, with the last two figures showing a reflection at the lower boundary of the domain.}
  \label{fig:ricker-wave-multi}
\end{figure*}

We placed a single Ricker wavelet source in the center of the domain. Figure~\ref{fig:ricker-wave-multi} illustrates that the shape and motion of the resulting waves behaved as expected, with a noticeably faster propagation speed at the deeper layers because of their higher wave velocities. Although absent at the upper boundary of the domain, we observed wave reflections in the lower boundary; they could be dealt with by employing absorbing boundary conditions \parencite{Engquist1977}. However, as the use of such conditions would require the PDE coefficients to be complex numbers, we do not address it here.


Although we performed all the experiments presented here with sufficiently small time steps to ensure consistent results, we made no rigorous investigations to establish the conditions for numerical stability. 

\section{Conclusions}
\label{sec:conclusions}

In this work, we proposed a memory reordering algorithm based on generalized Hilbert curves for use with the SEM to obtain improved memory efficiency and faster execution speed. We used this algorithm to implement a generic 2D wave equation solver using the SEM with unstructured meshes and an explicit time integration scheme.

The proposed algorithm derived from Hilbert curves presented the shortest compute time compared to three other commonly used approaches. This was observed not only with several mesh granularities, but also with large variations in element sizes. Simulation time reduced about 20\% when the element size variation was moderate (standard deviation varying between 0.29 and 0.58), and about 15\% when size variation was more significant (standard deviation varying between 1.55 and 3.79). This makes the proposed algorithm suitable to situations where specific regions of a mesh may demand particular refinements --- for instance, in stratified domains containing thin layers whose strong curvature requires relatively small elements.

In our experiments, memory reordering increased mesh preparation time by a value between 15\% and 20\%. In the complete simulation presented in Section~\ref{sec:realistic_case}, mesh preparation time was negligible within the total execution time, which is the case for most application scenarios with large unstructured meshes. Therefore, applying SFC-based memory reordering in all practical applications has a potential run time reduction of 25\%, as we have demonstrated, and hardly any downside.

We point out some suggestions for future work. Firstly, there is still room to improve the performance of the arithmetic code in Eqs.~\eqref{eq:u_change} and \eqref{eq:v_change}, in particular by precalculating more values. For instance, we can compute the multiplications between basis function derivatives and arrays $FK$/$BK$ before time integration starts, at the cost of an increase in memory consumption and additional memory to cache data traffic for the precalculated values. Secondly, the SEM formulation shown here can be readily extended to 3D by defining the continuous variable $u$ as a vector and modifying Eqs.~\eqref{eq:def_u_2} and \eqref{eq:def_v_2} accordingly. \textcite{Luo2006} provide local node locations and basis functions derived from Lobatto polynomials over tetrahedra that are well-suited for the SEM in the 3D case.

\section*{Declarations}

This research did not receive any specific grant from funding agencies in the public, commercial, or not-for-profit sectors.

\section*{Acknowledgments}

We would like to thank Adam Ellery of the University of Queensland for his instrumental help and advice regarding SEM and basis functions.

This research was supported by the High Performance Computing Center at UFRN (NPAD/UFRN).

\section*{Computer Code Availability}

The computer code produced in this research (version v20210104) is available for download at \url{https://gitlab.com/lappsufrn/shirley} under Apache License 2.0.

\nocite{*} 
\printbibliography[category=cited]

\end{document}